\documentclass[twocolumn]{aastex631}

\usepackage[english]{babel} % default (American) English hyphenation
\usepackage[utf8]{inputenc} % useful to type directly diacritic characters
\usepackage[T1]{fontenc}    % Use vector fonts, so it zooms properly in on-screen viewing software
\usepackage{ae,aecompl}
\usepackage{pgf,pgfarrows,pgfnodes,pgfautomata,pgfheaps}
\usepackage{graphicx}       % you can specify [driver] (e.g. [pdftex]) which is the dvi to postscript program
\usepackage{natbib}         % enables the use of \citep and others; see http://merkel.zoneo.net/Latex/natbib.php
\usepackage{url}            % enables the use of url web links
\usepackage{grffile}        % enables dots and underscores in .pdf filenames
\usepackage{mathtools}      % enables the use of various environments work, e.g., |align|, |dcases|, define symbols (see below), etc.
\usepackage{multirow}       % enables the use of \multirow{num}{*}{text} in deluxetable
\usepackage{xspace}         % enables the use of \xspace in defining macros
\usepackage{booktabs}       % enable \toprule, etc. in tables in proposals
\usepackage{txfonts}
\usepackage{amsmath,amssymb,amsxtra,amsfonts}   % to use pmatrix, etc.

\usepackage{color}         % using color package; put color boxes around text: \fcolorbox{frame colour}{background colour}{text}
\definecolor{gold}{rgb}{1,0.80,0}
\definecolor{orange}{rgb}{1,0.5,0}
\definecolor{midgray}{gray}{0.3}
\definecolor{lblue}{rgb}{0,0.2,0.6}
\definecolor{dgreen}{rgb}{0.1,0.6,0.3}
\definecolor{purple}{rgb}{0.5019607843137255,0.0,0.5019607843137255}

\renewcommand{\arcdeg}{\mbox{$^{\circ}$}\xspace}             % the unit of arcdeg
\newcommand{\be}{\begin{equation}}
\newcommand{\ee}{\end{equation}}

\newcommand{\ba}{\begin{align}}
\newcommand{\ea}{\end{align}}

\newcommand{\defeq}{\vcentcolon=}

\newcommand{\Msun}{\ensuremath{M_\odot}\xspace}

\newcommand{\Mstar}{\ensuremath{M_\ast}\xspace}

\newcommand{\oh}{\ensuremath{12+\log({\rm O/H})}\xspace}

\newcommand{\K}{\ensuremath{\rm K}\xspace}

\newcommand{\Hunit}{\ensuremath{\rm km~s^{-1}~Mpc^{-1}}\xspace}
\newcommand{\Funit}{\ensuremath{\rm erg~s^{-1}~cm^{-2}}\xspace}

\newcommand{\Ha}{\textrm{H}\ensuremath{\alpha}\xspace}
\newcommand{\Hb}{\textrm{H}\ensuremath{\beta}\xspace}
\newcommand{\Hg}{\textrm{H}\ensuremath{\gamma}\xspace}
\newcommand{\Hd}{\textrm{H}\ensuremath{\delta}\xspace}
\newcommand{\HII}{\textrm{H}\textsc{ii}\xspace}

\newcommand{\OII}{[\textrm{O}~\textsc{ii}]\xspace}
\newcommand{\OIII}{[\textrm{O}~\textsc{iii}]\xspace}

\newcommand{\NII}{[\textrm{N}~\textsc{ii}]\xspace}
\newcommand{\SII}{[\textrm{S}~\textsc{ii}]\xspace}
\newcommand{\NeIII}{[\textrm{Ne}~\textsc{iii}]\xspace}

\def\U{\ensuremath{U_{360}}\xspace}     % only for LBT U-band imaging
\def\B{\ensuremath{B_{435}}\xspace}
\def\V{\ensuremath{V_{606}}\xspace}
\def\I{\ensuremath{I_{814}}\xspace}
\def\Y{\ensuremath{Y_{105}}\xspace}
\def\J{\ensuremath{J_{125}}\xspace}

\def\H{\ensuremath{H_{160}}\xspace}

\newcommand{\emc}{\textsc{Emcee}\xspace}

\newcommand{\grzl}{\textsc{Grizli}\xspace}

\newcommand{\bagp}{\textsc{Bagpipes}\xspace}
\newcommand{\lmfit}{\textsc{LMfit}\xspace}

\newcommand{\jwst}{\textit{JWST}\xspace}

\newcommand{\glass}{\textit{GLASS}\xspace}

\def\ie{i.e.\xspace}
\def\eg{e.g.\xspace}

\def\p{{\rm prior}}

\newcommand{\Om} {\ensuremath{\Omega_{\rm{m}}}\xspace}

\newcommand{\Ol} {\ensuremath{\Omega_{\Lambda}}\xspace}

\newcommand{\n}{\ensuremath{{\nu}\rm}\xspace}

\begin{document}

\title{
Early results from GLASS-JWST. XXVII. The mass-metallicity relation in lensed field galaxies at cosmic noon with NIRISS\footnote{Based on observations acquired by the \jwst under the ERS program ID 1324 (PI T. Treu)}
}

\correspondingauthor{Xin Wang}
\email{xwang@ucas.ac.cn}

%%% Lead authors and main contributors
\author[0000-0002-1336-5100]{Xianlong He}
\affil{School of Astronomy and Space Science, University of Chinese Academy of Sciences (UCAS), Beijing 100049, China}
\affiliation{School of Physics and Technology, Wuhan University (WHU), Wuhan 430072, China}

\author[0000-0002-9373-3865]{Xin Wang}
\affil{School of Astronomy and Space Science, University of Chinese Academy of Sciences (UCAS), Beijing 100049, China}
\affil{National Astronomical Observatories, Chinese Academy of Sciences, Beijing 100101, China}
\affil{Institute for Frontiers in Astronomy and Astrophysics, Beijing Normal University, Beijing 102206, China}

\author[0000-0001-5860-3419]{Tucker Jones}
\affiliation{Department of Physics and Astronomy, University of California Davis, 1 Shields Avenue, Davis, CA 95616, USA}

\author[0000-0002-8460-0390]{Tommaso Treu}
\affiliation{Department of Physics and Astronomy, University of California, Los Angeles, 430 Portola Plaza, Los Angeles, CA 90095, USA}

\author[0000-0002-3254-9044]{K. Glazebrook}\affiliation{Centre for Astrophysics and Supercomputing, Swinburne University of Technology, PO Box 218, Hawthorn, VIC 3122, Australia}

\author[0000-0001-6919-1237]{Matthew A. Malkan}
\affiliation{Department of Physics and Astronomy, University of California, Los Angeles, 430 Portola Plaza, Los Angeles, CA 90095, USA}

\author[0000-0003-0980-1499]{Benedetta Vulcani}
\affiliation{INAF Osservatorio Astronomico di Padova, vicolo dell'Osservatorio 5, 35122 Padova, Italy}

\author[0000-0002-8632-6049]{Benjamin Metha}
\affiliation{School of Physics, University of Melbourne, Parkville 3010, VIC, Australia}
\affiliation{ARC Centre of Excellence for All Sky Astrophysics in 3 Dimensions (ASTRO 3D), Australia} 
\affiliation{Department of Physics and Astronomy, University of California, Los Angeles, 430 Portola Plaza, Los Angeles, CA 90095, USA}

%%% NIRISS builders (see GLASS publication policy doc)

\author[0000-0001-5984-0395]{Maru\v{s}a Brada\v{c}}
\affiliation{University of Ljubljana, Department of Mathematics and Physics, Jadranska ulica 19, SI-1000 Ljubljana, Slovenia}
\affiliation{Department of Physics and Astronomy, University of California Davis, 1 Shields Avenue, Davis, CA 95616, USA}

\author[0000-0003-2680-005X]{Gabriel Brammer}
\affiliation{Cosmic Dawn Center (DAWN), Denmark}
\affiliation{Niels Bohr Institute, University of Copenhagen, Jagtvej 128, DK-2200 Copenhagen N, Denmark}

\author[0000-0002-4140-1367]{Guido Roberts-Borsani}
\affiliation{Department of Physics and Astronomy, University of California, Los Angeles, 430 Portola Plaza, Los Angeles, CA 90095, USA}

\author[0000-0002-6338-7295]{Victoria Strait}
\affiliation{Cosmic Dawn Center (DAWN), Denmark}
\affiliation{Niels Bohr Institute, University of Copenhagen, Jagtvej 128, DK-2200 Copenhagen N, Denmark}

%%% NIRCam builders (see GLASS publication policy doc)

\author[0000-0002-2667-5482]{Andrea Bonchi}
\affiliation{Space Science Data Center, Italian Space Agency, via del Politecnico, 00133, Roma, Italy}
\affiliation{INAF Osservatorio Astronomico di Roma, Via Frascati 33, 00078 Monteporzio Catone, Rome, Italy}

\author[0000-0001-9875-8263]{Marco Castellano}
\affiliation{INAF Osservatorio Astronomico di Roma, Via Frascati 33, 00078 Monteporzio Catone, Rome, Italy}

\author[0000-0003-3820-2823]{Adriano Fontana}
\affiliation{INAF Osservatorio Astronomico di Roma, Via Frascati 33, 00078 Monteporzio Catone, Rome, Italy}

\author[0000-0002-3407-1785]{Charlotte Mason}
\affiliation{Cosmic Dawn Center (DAWN), Denmark}
\affiliation{Niels Bohr Institute, University of Copenhagen, Jagtvej 128, DK-2200 Copenhagen N, Denmark}

\author[0000-0001-6870-8900]{Emiliano Merlin}
\affiliation{INAF Osservatorio Astronomico di Roma, Via Frascati 33, 00078 Monteporzio Catone, Rome, Italy}

\author[0000-0002-8512-1404]{Takahiro Morishita}
\affiliation{IPAC, California Institute of Technology, MC 314-6, 1200 E. California Boulevard, Pasadena, CA 91125, USA}

\author[0000-0002-7409-8114]{Diego Paris}
\affiliation{INAF Osservatorio Astronomico di Roma, Via Frascati 33, 00078 Monteporzio Catone, Rome, Italy}

\author[0000-0002-9334-8705]{Paola Santini}
\affiliation{INAF Osservatorio Astronomico di Roma, Via Frascati 33, 00078 Monteporzio Catone, Rome, Italy}

\author[0000-0001-9391-305X]{Michele Trenti}
\affiliation{School of Physics, University of Melbourne, Parkville 3010, VIC, Australia}
\affiliation{ARC Centre of Excellence for All Sky Astrophysics in 3 Dimensions (ASTRO 3D), Australia}

%%% Additional contributors in alphabetical order
\author[0000-0003-4109-304X]{Kristan Boyett}
\affiliation{School of Physics, University of Melbourne, Parkville 3010, VIC, Australia}
\affiliation{ARC Centre of Excellence for All Sky Astrophysics in 3 Dimensions (ASTRO 3D), Australia}

\author[0000-0002-3247-5321]{K. Grasha}
\altaffiliation{ARC DECRA Fellow}
\affiliation{Research School of Astronomy and Astrophysics, Australian National University, Canberra, ACT 2611, Australia}   
\affiliation{ARC Centre of Excellence for All Sky Astrophysics in 3 Dimensions (ASTRO 3D), Australia}  
\affiliation{Visiting Fellow, Harvard-Smithsonian Center for Astrophysics, 60 Garden Street, Cambridge, MA 02138, USA}   

% \collaboration{21}{(+ others)}

\begin{abstract}

We present a measurement of the mass-metallicity relation (MZR) at cosmic noon, using the JWST near-infrared wide-field slitless spectroscopy obtained by the GLASS-JWST Early Release Science program. By combining the power of JWST and the lensing magnification by the foreground cluster A2744, we extend the measurements of the MZR to the dwarf mass regime at high redshifts. A sample of 50 galaxies with several emission lines is identified across two wide redshift ranges of $z=1.8-2.3$ and $2.6-3.4$ in the stellar mass range of $\log{(M_*/M_\odot)}\in [6.9, 10.0]$. The observed slope of MZR is $0.223 \pm 0.017$ and $0.294 \pm 0.010$ at these two redshift ranges, respectively, consistent with the slopes measured in field galaxies with higher masses. In addition, we assess the impact of the morphological broadening on emission line measurement by comparing two methods of using 2D forward modeling and line profile fitting to 1D extracted spectra. We show that ignoring the morphological broadening effect when deriving line fluxes from grism spectra results in a systematic reduction of flux by $\sim30\%$ on average. This discrepancy appears to affect all the lines and thus does not lead to significant changes in flux ratio and metallicity measurements. This assessment of the morphological broadening effect using JWST data presents, for the first time, an important guideline for future work deriving galaxy line fluxes from wide-field slitless spectroscopy, such as Euclid, Roman, and the Chinese Space Station Telescope.

\end{abstract}

% Classical Novae (251) --- Ultraviolet astronomy(1736) --- History of astronomy(1868) --- Interdisciplinary astronomy(804)


\section{Introduction} \label{sec:intro}

Nearly all elements heavier than Helium (referred to as metals in astronomy) are synthesized by stellar nuclear reactions, making them a good tracer of star formation activity across cosmic time.
Star formation rate (SFR) and metal enrichment peak at the “Cosmic Noon” epoch $z\sim2$ \citep[Fig.9]{Madau:2014ARA&A..52..415M}, confirmed by a census of deep surveys with Hubble Space Telescope (HST), the Sloan Digital Sky Survey (SDSS), and other facilities.
Metals are thought to be expelled into the interstellar/intergalactic medium (ISM/IGM) by stellar explosions such as supernovae and stellar winds.
The cumulative history of the baryonic mass assembly, e.g., star formation, gas accretion, mergers, feedback, and galactic winds, altogether governs the total amount of metals remaining in gas \citep{Finlator:2008MNRAS.385.2181F, Dave:2012MNRAS.421...98D, Lilly:2013ko, Dekel:2014jm, Peng:2014hn}. 
Therefore, the elemental abundances provide a crucial diagnostic of the past history of star formation and complex gas movements driven by galactic feedback and tidal interactions \citep{Lilly:2013ko, Maiolino:2019vq}.
Since detailed abundances are not directly measurable at extragalactic distances, the relative oxygen abundance (number density) compared to hydrogen in ionized gaseous nebulae (reported as \oh), is often chosen as the observational proxy of \emph{metallicity} for simplicity.

Several scaling relations have been established, characterizing the tight correlations between various physical properties of star-forming galaxies,  \eg, stellar mass (\Mstar), metallicity $Z$, SFR, luminosity, size, and morphology \citep[see][for recent reviews]{Kewley:2019kf, Maiolino:2019vq}.
Metallicity abundance evolution was found to exhibit a strong correlation with mass during galaxy evolution history \citep{ Dave:2011MNRAS.416.1354D,Lu:2015vd}.
The mass–metallicity relation (MZR), has been quantitatively established in the past two decades in both the local \citep[mainly from SDSS]{Tremonti:2004ed, Zahid:2012fp, Andrews:2013dn}, and the distant universe out to $z\sim3$ \citep{Erb:2006kn,Maiolino:2008A&A...488..463M,Zahid:2011bb,Henry:2013gx, Henry:2021ju,Sanders:2015gk,Sanders:2021ga}.
Recently, the launch of JWST has enabled the measurement of the MZR out to $z\sim8$ \citep[e.g.,][]{Arellano-Cordova:2022ApJL, Schaerer:2022A&A,Trump:2023ApJ, Rhoads:2023ApJL,Curti:2023arXiv230408516C,Curti:2023MNRAS,Nakajima:2023arXiv230112825N,Sanders:2023arXiv230308149S, Matthee:2023ApJ}. 
The slope of the MZR is sensitive to the properties of outflows (e.g., mass loading factor, gas outflow velocity), which are a crucial ingredient to galaxy evolution models \citep[see][]{Dave:2012MNRAS.421...98D, Lu:2015kh,Henry:2021ju}.
The MZR slope has also been used to reveal trends in how the star formation efficiency and galaxy gas mass fraction depend on stellar mass \citep{Baldry:2008hm,Zahid:2014ia}.
\citet{Mannucci:2010MNRAS.408.2115M} first suggested a so-called fundamental metallicity relation (FMR), 
which aims to explain the scatter and redshift evolution of the MZR by introducing the SFR as an additional variable, creating a 3-parameter scaling relation.
The FMR has a small intrinsic scatter of $\sim0.05$ dex in metallicity,
making it possible to trace the metal production rates in stellar within cosmological time \citep{Finlator:2008MNRAS.385.2181F}.
Moreover, spatially resolved chemical information encoded by the metallicity radial gradients \citep{Jones:2015AJ....149..107J,Wang:2016um,Wang:2019cf,Wang:2020bp,Wang:2022ApJ...926...70W,Franchetto:2021ApJ}, is a sensitive probes of baryonic assembly and the complex gas flows driven by both galactic feedback and tidal interactions.

The Near-infrared Imager and Slitless Spectrograph \citep[NIRISS;][]{Willott:2022PASP..134b5002W} onboard the JWST now enables a tremendous leap forward with its superior sensitivity, angular resolution, and longer wavelength coverage compared to HST/WFC3. This allows metallicity measurements with better precision in galaxies with lower stellar mass at the cosmic noon epoch $1<z<3$.
Similar measurements have been done using data from NIRSpec gratings \citep[e.g.,][]{Shapley:2023ApJL,Curti:2023MNRAS}, NIRSpec prism \citep[][]{Langeroodi:2023ApJ}, NIRCam WFSS \citep{Matthee:2023ApJ}, and NIRISS \citep{Li:2022arXiv221101382L}.
This paper takes advantage of the deep NIRISS spectroscopy acquired by the Early Release Science (ERS) program \glass-\jwst \citep[ID ERS-1324\footnote{\url{https://www.stsci.edu/jwst/science-execution/approved-programs/dd-ers/program-1324}};][]{Treu:2022ApJ...935..110T} in the field of the galaxy cluster Abell 2744 (A2744).
By exploiting the gravitational lensing magnification produced by the foreground A2744 cluster, we are able to extend the measurement of the MZR down to $10^7$ solar mass $M_\odot$.

In this paper, we present a measurement of the MZR using the NIRISS and NIRCam data from a sample of 50 lensed field galaxies in a low mass range at $z\sim2-3$.
In Sect.~\ref{sec:data}, we describe the data acquisition and galaxy sample analyzed in this work.
In Sect.~\ref{sec:measure}, we demonstrate our method to extract metallicity and stellar mass for both individual galaxies and their stacked spectrum.
The main goal of this work is to present our MZR measurements in Fig.~\ref{fig:mzr}.
We discuss the results in Sect.~\ref{sec:result} and summarize the main conclusions in Sect.~\ref{sec:conclude}.
The AB magnitude system, the standard concordance cosmology ($\Om=0.3, \Ol=0.7$, $H_0=70\,\Hunit$), and the \citet{Chabrier:2003ki} Initial Mass Function (IMF) are adopted.  
The metallic lines are denoted in the following manner, if presented without wavelength: $\OII~\lambda\lambda3727,3730\defeq\OII,\NeIII~\lambda3869\defeq\NeIII,$
%\xl{$\Hd3869\defeq\Hd$ not used?},
$ \Hg~\lambda4342\defeq\Hg, \Hb~\lambda4863\defeq\Hb, \OIII~\lambda5008\defeq\OIII, \Ha~\lambda6564\defeq\Ha, \SII~\lambda\lambda6716,6731\defeq\SII $.

\section{Observation Data} \label{sec:data}

We use the joint JWST NIRISS and NIRCam data targeting the A2744 lensing field cluster. 
The NIRISS data are used to estimate the metallicity through modeling of emission line flux ratios, while the NIRCam data are used to calculate the stellar mass through Spectral Energy Distribution (SED) Fitting.

The spectroscopy data from JWST/NIRISS of GLASS-ERS (program DD-ERS-1324, PI: T. Treu),  with the observing strategy described by \citet{Treu:2022ApJ...935..110T}, is reduced in Paper I \citep{Roberts:2022ApJ...938L..13R}. 
Briefly, the core of the A2744 cluster ($130"\times 130"$) was observed for $\sim18.1$ hr with NIRISS wide-field slitless spectroscopy and direct imaging for $\sim2.36$ hr in three filters (F115W, F150W, and F200W)\footnote{see the official documentation for more information: \url{https://jwst-docs.stsci.edu/jwst-near-infrared-imager-and-slitless-spectrograph/niriss-observing-modes/niriss-wide-field-slitless-spectroscopy}} on June 28-29, 2022 and July 07, 2023.
The total exposure times for the majority of sources in each of these three bands amount to 5.4, 5.7, 2.9 hours (as detailed in Fig.~\ref{fig:indivi_spec_example}).
This provides low-resolution $R:=\lambda/\Delta\lambda\sim150$ spectra of all objects in the field of view with continuous wavelength coverage from $\lambda\in [1.0,2.2]$ $\mu$m.
This includes the strong rest-frame optical emission lines \OII, \NeIII, \Hg, \Hb,  \OIII at $z\in[1.8, 3.4]$, and \Ha, \SII at $z\in[1.8, 2.3]$\footnote{\citet{Li:2022arXiv221101382L} have developed an interactive website to visualize the emission lines covered by each filter at different redshifts: \url{https://preview.lmytime.com/jwstfilter}}.
Spectra are taken at two orthogonal dispersion angles (using the GR150C and GR150R grism elements), which helps to minimize the effects of contamination by overlapping spectral traces.

The photometric data of the A2744 cluster we used are the publicly released NIRCam images \citep{Paris:2023arXiv230102179P}, coming from three programs: GLASS-JWST (PI Treu), UNCOVER (PIs Bezanson and Labbé), and DDT-2756 (PI Chen).
It is an F444W-detected multi-band catalog, including all NIRCam and available HST data. 
All reduced images in 8 JWST/NIRCam bands (F090W, F115W, F150W, F200W, F277W, F356W, F410M, F444W), 4 HST/ACS-WFC bands (F435W, F606W, F775W, F814W) and 4 HST/WFC3-IR bands (F105W, F125W, F140W, F160W)\footnote{see the repository of the Spanish Virtual Observatory for more Filter information: \url{http://svo2.cab.inta-csic.es/theory/fps/}} are used if available.
This photometric data, with an observed-frame wavelength coverage of  $0.4-5$ $\mu$m at redshift $z\in[1.8, 3.4]$, enables very good stellar mass estimates by sampling the full rest-UV to near-IR SEDs. 
We also use the half-light radius $r
_{50}$ of this catalogue in Sect.~\ref{subsec:lineflux}. 
The half-light radius $r
_{50}$ is computed by \textsc{SExtractor}\xspace in the F444W band in units of pixel (the effective radius FLUX\_RADIUS in \textsc{SExtrator}).

\section{Measurements}\label{sec:measure}

In this section, we present the measurements of the physical properties derived from spectroscopy and photometry, with the result of 50 individual galaxies shown in Tab. \ref{tab:indi}.

Quantities (e.g., the stellar mass $M_*$ and SFR) that are derived from a single flux must be corrected for the modest gravitational lensing magnification by the foreground A2744 cluster. 
But properties that are derived from flux ratio (e.g., metallicity $Z$) or other observed quantities, are independent of lensing magnification.
We adopt our latest high-precision, JWST-based lensing model \citep{Bergamini:2023arXiv230310210B, Bergamini:2023A&A...670A..60B} to estimate the lensing magnification $\mu$.
We do not consider the uncertainty of $\mu$ because the relative error is only  $\sim2.3\%$.
The median estimate of $\mu$ is consistent but more precise with the calculation derived from the public Hubble Frontier Fields (HFF) lensing tool\footnote{\url{https://archive.stsci.edu/prepds/frontier/lensmodels/}} \citep{Lotz:2017ApJ...837...97L} using \textit{Sharon \& Johnson} version \citep{Johnson:2014cf} and the \textit{CATS} version \citep{Jauzac:2015df} computed by \textsc{Lenstool} software\footnote{\url{https://lenstools.readthedocs.io/en/latest/}} \citep{Petri:2016A&C....17...73P}.

\subsection{Grism Redshift and Emission-line Flux}\label{subsec:gri}

\begin{figure*}
    \centering
    \includegraphics[width=\textwidth]{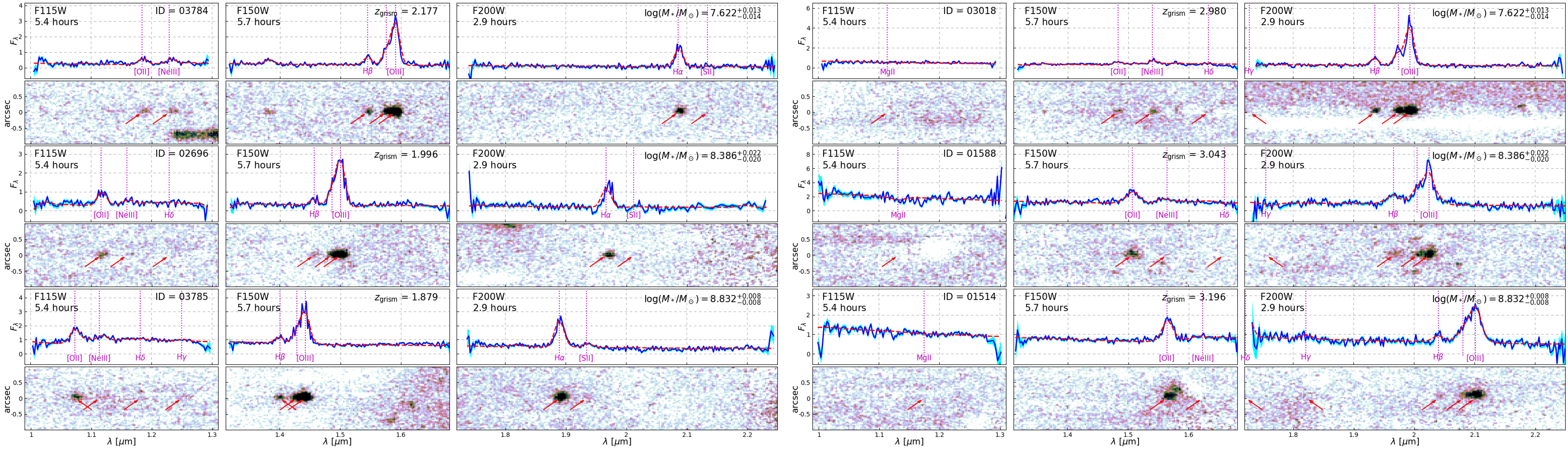}
    \caption{1D/2D spectra of six galaxies in our sample. \textit{Left:} three examples at $z=1.8-2.3$. 
    The forward-modeled spectra, optimally extracted 1D observed flux $F_\lambda$ (in units of [10$^{-19}$ erg/s/cm$^2$/$\AA$]), and its $1\sigma$ uncertainty, are represented by the red, blue solid lines and the cyan shaded bands, respectively. 
    The 2D grism spectra covered in three filters (F115W, F150W, F200W) are continuum-subtracted.
    \textit{Right:} same as in \textit{Left} but at $z=2.6-3.4$.
    }
    \label{fig:indivi_spec_example}
\end{figure*}

We utilize the Grism Redshift and Line Analysis software \textsc{Grizli} \citep{brammer_gabriel_2023_7712834} to reduce NIRISS data using the standard JWST pipeline (version 1.11.1) and the latest reference file (under the \textsc{jwst\_1100.pmap} context). The detailed procedures are largely described in \cite{RobertsBorsani:2022ApJ...938L..13R}. 
Briefly, \grzl analyzes the paired direct imaging and grism exposures through forward modeling, and yields contamination subtracted 1D \& 2D grism spectra, along with the best-fit spectroscopic redshifts. 

For each source, the one dimensional (1D) spectrum is constructed using a linear superposition of a spectra from a library consisting of four sets of empirical continuum spectra covering a range of stellar population ages \citep{Brammer:2008gn,Erb:2010iy,Muzzin:2013is,Conroy:2012bl} and Gaussian-shaped nebular emission lines at the observed wavelengths given by the source redshift.
The intrinsic 1D spectrum and the spatial distribution of flux measured in the paired direct image are utilized to generate a 2D model spectrum based on the grism sensitivity and dispersion function, similar to the ``fluxcube'' model produced by the aXe software \citep{2009PASP..121...59K}.
This 2D forward-modeled spectrum is then compared to the observation by \grzl and a global $\chi^2$ calculation is performed to determine the best-fit superposition coefficients for both the continuum templates and Gaussian amplitudes, the latter of which correspond to the best-fit emission line fluxes.
In this way, our 2D forward modeling practice not only determines the source redshift, but also measures the emission line fluxes, taking into account the morphological broadening effect.
We refer the interested readers to Appendix A of \citet{Wang:2019cf}, for the full descriptions of the redshift fitting procedure.

We obtain a parent sample of 4756 sources with F150W apparent magnitudes between [18, 32] ABmag (the $5\sigma$ depth is 28.7 according to \citet{Treu:2022ApJ...935..110T}), on which our \grzl analyses result in meaningful redshift constraints.
Several goodness-of-fit criteria are implemented to ensure the reliability of our redshift fit:
a reduced chi-square close to 1 $(\chi^2<2.2)$,  a sharply peaked posterior of the redshift $\left(\Delta z\right)_{\rm posterior}/(1+z_{\rm peak})<0.002$, high evidence of Bayesian information criterion compared to polynomials $(\mathrm{BIC}>100)$.
As a result, there are 348 sources in the redshift range $z\in[0.05, 10]$, with secure grism redshift measurements according to the above joint selection criteria.
A total of 86 sources with secure grism redshifts are at redshifts $z\in[1.8, 2.3]\cup [2.6, 3.4]$, ensuring that the slitless spectra cover several emission lines:\OII, \NeIII, \Hd, \Hg, \Hb, \OIII (also \Ha, \SII for the former zone), with high sensitivity for our 3 NIRISS filters (F115W, F150W, and F200W). 
However, 6/86 sources of our NIRISS spectroscopy catalog do not match entries in the NIRCam photometric catalog \cite{Paris:2023arXiv230102179P} within 0.7 arcsecs ($5\times$ PSF).

The fluxes of the intrinsic nebular emission lines (\OII, \NeIII, \Hd, \Hg, \Hb, \OIII, \Ha, and \SII, the same as in \citealt{Henry:2021ju}) are 2D forward modeled by \textsc{Grizli} as output. 
There are 57 sources with \Hb detection, to ensure the reliable measurement of SFR.
No other emission line criteria (e.g., SNR \OIII) are used for selection, to avoid potential metallicity bias.
Then we visually inspect the 1D spectra of each galaxy individually, excluding 7 of those that are heavily contaminated. 
The 50 galaxies showing prominent nebular emission features, with 0 possible AGNs exclusion in Sect.~\ref{subsec:AGN}, will make up the final sample presented in Tab. \ref{tab:indi}.
A 'textbook case' of our samples (ID: 05184 in Tab.\ref{tab:indi}) has been carefully studied through spatial mapping in our recent work \citep{Wang:2022ApJ...938L..16W}. 
We show as an example 1D/2D spectra for six galaxies in our sample in Fig.~\ref{fig:indivi_spec_example}, annotated with their exposure times, best-fit grism redshifts, and stellar masses (which will be discussed in Sect.~\ref{subsec:mass}).

Since the 1D grism spectra are extracted by \grzl simultaneously, it allows us to directly fit it using several 1D Gaussian profiles to obtain line fluxes and errors, as detailed in Sect.~\ref{subsec:stack}.
But we still use the previous 2D flux other than 1D as our default result for subsequent calculations.
The comparison of the line flux measurements between this 1D line profile fitting and the 2D \grzl forward-modeling procedure, is discussed in Sect.~\ref{subsec:lineflux}. 

\subsection{Gas-phase metallicity and Star Formation Rate} 
\label{subsec:metal}

We use these observed line flux $(f_i^\text{o}, \sigma_i^\text{o})$ to simultaneously estimate 3 parameters: jointly metallicity, nebular dust extinction, and de-reddened $\Hb$ line flux (\oh, $A_v$, $f_{\Hb}$).
We follow the previous series of work \citep{Jones:2015AJ....149..107J, Wang:2016um,Wang:2019cf,Wang:2020bp,Wang:2022ApJ...926...70W}, by constructing a Bayesian inference method that uses multiple calibration relations to jointly constrain metallicity \oh, and ($A_v$, $f_{\Hb}$) simultaneously.
Our method is more reliable than the conventional way of turning line flux ratios into metallicities, since it takes into account the intrinsic scatter in strong-line O/H calibrations ($\sigma_{R_i}$ in Eq.~\ref{eq:chi_square} ).
And it combines multiple line flux measurements and properly marginalizes over the dust extinction correction.
It also emphasizes bright lines (\eg, \OII, \OIII) with high signal-to-noise ratios (SNRs) and marginalizes faint lines (\eg, \Hb) or even non-detection lines with low SNRs quantitatively, (\ie, by assigning weights to each line according to its SNR in the likelihood function).

The Markov Chain Monte Carlo (MCMC) sampler \emc software \citep{ForemanMackey:2013io} is employed to sample the likelihood profile $\mathcal{L}\propto\exp{(-\chi^2/2)}$ with:
\begin{equation}
    \chi^2 := \sum_{i}^\mathrm{EL} \frac{\left(f_{i}-R_i \cdot {f_{\mathrm{H}\beta}}\right)^2}{\left(\sigma_{i}\right)^2+ \left(\sigma_{R_i}\right)^2 \cdot {f_{\mathrm{H}\beta}} ^2 }, \quad R_i:=\frac{f_i}{f_{H\beta}}. 
    \label{eq:chi_square}
\end{equation}
Here the summation $i$ includes all emission lines,  with their intrinsic scatters $\sigma_{R_i} := \sigma_i^{\text{cal}} \cdot R_i \cdot \ln{10}$. 
The inherent flux and uncertainty $(f_i, \sigma_i)$ for each line, are corrected from observation $(f_i^\text{o}, \sigma_i^\text{o})$ for dust attenuation by parameter $A_v$ using the \citet{Calzetti:2000iy} extinction law. 
$R_i$ refers to the line flux ratio, which is empirically calibrated by a polynomial as a function of metallicity: $\log{R} = \sum_{j=0}^n c_{j} \cdot (x)^j, x:=\oh$,
where $(x)^j$ means $j$th power of $x$, with the coefficients summarized in Tab.~\ref{tab:coef}.
For flux ratio calibrations that do not use \Hb as the denominator (e.g.,  \NeIII/\OIII), the terms $f_{\Hb}$ in Eq.\ref{eq:chi_square} need to be replaced by the corresponding lines (e.g., $f_{\OIII}$). And one more term of uncertainty (e.g., $\sigma_{\text{O3}}^2\cdot R_\text{Ne3}^2$) needs to be added to the denominator of $\chi^2$.

% = = = = = = = = = = = = = = = = = = = = = = = = = = = = = = = = = = = = = = = = = =
% Include this table with \input{filename.tex}
% To rotate in emulateapj do: \begin{turnpage}\input{filename.tex}\end{turnpage}
% To display it on multiple pages do: \LongTables\input{filename.tex}
% - - - - - - - - - - - - - - - - - - - - - - - - - - - - - - - - - - - - - - - - - -
%{
%\tabletypesize{\scriptsize}
%\tabcolsep=2pt
\begin{deluxetable*}{lcccccccccccc}
    \tablecolumns{9}
    \tablewidth{0pt}
    \tablecaption{Coefficients for the emission line flux ratio diagnostics used in this work.}
% - - - - - - - - - - - - - - - - - - - - - - - - - - - - - - - - - - - - - - - - - -
\tablehead{
    \colhead{Diagnostic $R$ and Notation} &
    \colhead{$c_0$} &
    \colhead{$c_1$} &
    \colhead{$c_2$} &
    \colhead{$c_3$} &
    \colhead{$c_4$} &
    \colhead{$c_5$} &
    \colhead{$\sigma^\text{cal}$} &
    \colhead{ref}
}
%---------------------------------------------------------------
\startdata
    % \multicolumn{9}{c}{Strong line calibrations of \citet[][B18]{Bian:2018km}\tablenotemark{a}}   \\
    \noalign{\smallskip}
    $\text{O}_3:=$ \OIII/\Hb   &   43.9836    &   -21.6211   &   3.4277  &   -0.1747 & \nodata & \nodata & 0.05 & \multirow{2}{*}{\citet[][B18]{Bian:2018km} }\\
    $\text{O}_2:=$ \OII/\Hb    &   78.9068    &   -45.2533   &   7.4311  &   -0.3758 & \nodata & \nodata & 0.05 & \\
    \noalign{\smallskip}\hline\noalign{\smallskip}
    % \multicolumn{9}{c}{Balmer decrement}   \\
    \noalign{\smallskip}
    \Ha/\Hb   &   0.45637   &   \nodata   &   \nodata  &   \nodata &\nodata& \nodata& 0.00 & \multirow{2}{*}{Balmer decrement} \\
    \Hg/\Hb   &   -0.32790    &   \nodata   &   \nodata  &   \nodata &\nodata&\nodata& 0.00 & \\
    % \Hd &&&&&&&& \xl{need? no}\\
    % \noalign{\smallskip}\hline\noalign{\smallskip}
    $\text{Ne}_3\text{O}_3:=$ \NeIII/\OIII & -1.11420 &\nodata&\nodata&\nodata&\nodata&\nodata& 0.04 & \citet{Jones:2015ApJ...813..126J}\\
    \noalign{\smallskip}\hline\noalign{\smallskip}
    % \multicolumn{5}{c}{Strong line calibrations of \citet[][C17]{Curti:2017fn}\tablenotemark{b}}   \\
    % \multicolumn{9}{c}{calibrations of \tocite{Jones2015}}\\
    \noalign{\smallskip}
    & -0.54571 & 0.45730 & -0.82269 & -0.02839 & 0.59396 & 0.34258 & best \\
    $\text{S}_2:=$ \SII/\Ha & -0.43974 & 0.34034 & -0.62850 & -0.07077 & 0.47147 & 0.31767 &  upper & \citet{Jones:2015ApJ...813..126J}\tablenotemark{b} \\
    & -0.65464 & 0.58976 & -1.06047 & 0.01979 & 0.75382 & 0.37766 & lower\\
\enddata
% - - - - - - - - - - - - - - - - - - - - - - - - - - - - - - - - - - - - - - - - - -
    % \tablecomments{The empirical flux ratios are computed in the polynomial functional form of $\log{R} = \sum_{i} c_i\cdot x^i$, where $x=\oh-8.69$ for the \citet{Curti:2017fn} calibrations and $x=\oh$ for the \citet{Bian:2018km} ones.  }
    \tablenotetext{a}{We note that the \OIII/\Hb calibration reported in \citet{Bian:2018km} in fact refers to the flux ratio between \OIII4960,5008 and \Hb, \ie, a factor of (2.98/3.98) is needed \citep[following][]{Storey:2000jd} when we use the doublets let to calibrate pure \OIII5008.}
    % \tablenotetext{b}{The intrinsic scatter of 0.09(0.11) dex for the \OIII/\Hb(\OII/\Hb) calibrations, as reported in \citet{Curti:2017fn}, have been included in the metallicity inference.}
    \tablenotetext{b}{the line flux ratio $R_{\SII}$ is calibrated by polnomial with coefficients given by the 'best' row, and the uncertainty $\sigma_{\SII}$ is given by the `upper' and `lower' row, where the metallicity $x$ is relative to solar $x:=\oh-8.69$.}
\label{tab:coef}
\end{deluxetable*}
%}

A wide range of strong line calibrations between line flux ratio and metallicity has been established \citep[see Appendix C in][for a summary]{Wang:2019cf} \citep[also see][for recent reviews]{Maiolino:2019vq, Kewley:2019kf}.
Different choices can result in offsets as high as 0.7 dex \citep[see \eg,][]{Kewley:2008be}.
In this work, we adopt mainly the diagnostics group "$\text{O}_3-\text{O}_2$" of calibrations prescribed by \citet[][hereafter B18]{Bian:2018km}, for comparison with \citet{Sanders:2021ga,Wang:2022ApJ...926...70W}.
The purely empirical calibrations in \citet[][B18]{Bian:2018km} are based on a sample of local analogs of high-$z$ galaxies according to the location on the BPT diagram, with the notations and coefficients summarized in Tab.~\ref{tab:coef}.

These calibrations are recommended for the metallicity range of $7.8<12+\text{log(O/H)}$, which is appropriate for our sample that does not reach metallicities as low as those found at higher redshift \cite{Curti:2023arXiv230408516C,Heintz:2023NatAs}. 
As a sanity check, we computed metallicities using the calibrations from \citet{Sanders:2023arXiv230308149S}, and indeed we do not find galaxies with metallicities significantly lower than 7.8.
In order to make complete use of emission lines of spectra, we also collect $\mathrm{Ne_3O_3},\mathrm{S_2}$ diagnostics at the same time, even though the corresponding line fluxes are not so strong for our sample.
We have tested that if they are removed, they do not significantly affect the metallicity estimation, which is dominated by the first 2 diagnostics $\mathrm{O}_3,\mathrm{O}_2$ in B18 and 2 Balmer decrements.
We adopt the intrinsic Balmer decrement flux ratios assuming Case B recombination with $T_{\rm e}\sim10,000 {\rm K}$.
We neglect the line-blending effect, since they are likely small in most cases \citep[see Fig.~4 and Append.~C in][for more information]{Henry:2021ju}.
This Bayesian method is used to derive properties (\oh, $A_v$, $f_{\Hb}$) of galaxies both from our individual spectra sample here and from the stacked spectra presented in Sect.\ref{subsec:stack}.

From the de-reddened \Hb flux $f_{\Hb}$, we estimate the instantaneous SFR of our sample galaxies, based on Balmer line luminosities. 
This approach provides a valuable proxy of the ongoing star formation on a time scale of $\sim$10Myr, highly relevant for galaxies displaying strong nebular emission lines.
Assuming the \citet{Kennicutt:1998ki} calibration and the Balmer decrement ratio of $\Ha/\Hb=2.86$ from the case B recombination for typical \HII regions, we calculate:
\begin{equation}
    \mathrm{SFR} = 4.65\times10^{-42} \frac{L(\Hb)}{\mathrm{[erg\,s^{-1}]}}\times 2.86\,[\Msun\,\mathrm{yr}^{-1}],    
\end{equation}
suitable for the \citet{Chabrier:2003ki} initial mass function.
The total luminosity  $L(\Hb)=4\pi D_L^2(z) \cdot f_{\Hb}$ is corrected for lensing magnification according to \citet{Bergamini:2023arXiv230310210B}.
The corrected SFR values are given in Tab.~\ref{tab:indi}.

\begin{figure}
    \centering
    \includegraphics[width=\columnwidth]{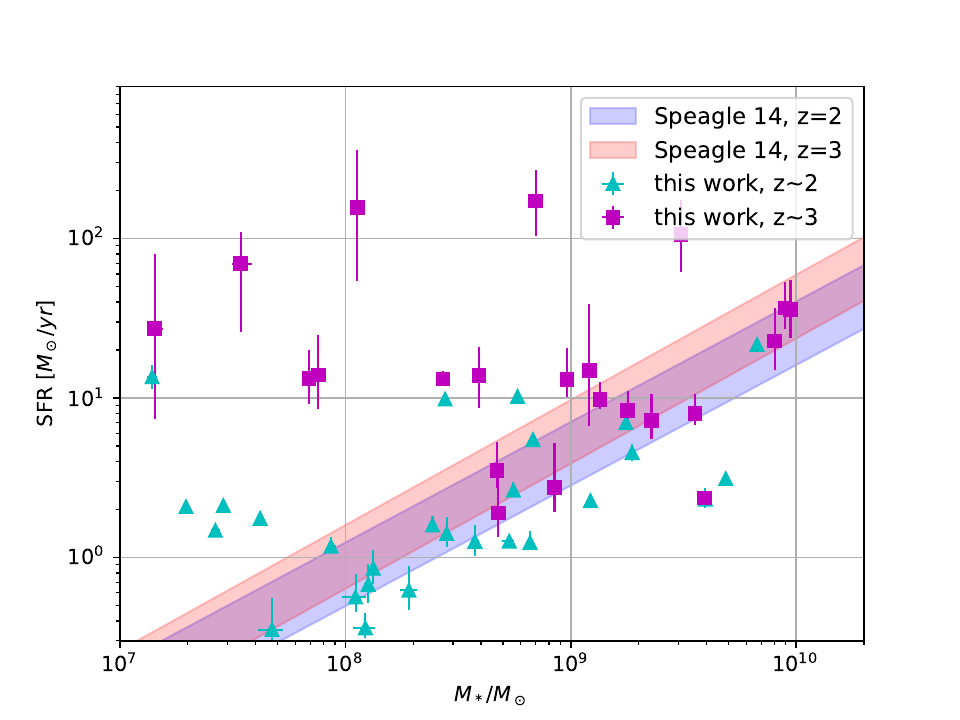}
    \caption{SFR-$M_*$ relation for our galaxy sample, where the low and high redshift individual measurements are marked in cyan triangles and magenta squares. As a comparison, we also show the star-forming main sequence fitted by \citet{Speagle:2014dd} with $\pm0.2$ dex scatters. \citet{Sanders:2021ga} gives results fairly close to their extrapolated best fit out to $\log(M_*/M_\odot)=9$.}
    \label{fig:sfr}
\end{figure}

\subsection{Stellar mass and Lensing magnification}\label{subsec:mass}

In this section, we fit broad-band photometry to obtain stellar mass $M_*$ of target galaxies through SED fitting.
We directly use the combined photometric catalog\footnote{\url{https://glass.astro.ucla.edu/ers/external_data.html}} released by the GLASS-JWST team \citep{Paris:2023arXiv230102179P}.
The photometric fluxes measured within $2\times$ PSF FWHM apertures of all 16 bands are included if available. 
We match 2983/4756 galaxies of our NIRISS spectroscopy catalog in Sect.\ref{subsec:gri} to the 24389 galaxies of the NIRCam photometric catalog with on-sky distances (d2d) lower than 0.7 arcsecs ($5\times$ FWHM in the
F444W band, conservatively).
As done in Sect. \ref{subsec:gri}, the final selected sample of 50 galaxies yields accurate d2d match ($<0.14$ arcsecs, around the angular resolution of JWST/NIRISS), and visually cross-matching with the NIRCam image further validates our sources.

To estimate the stellar masses \Mstar of our sample galaxies, we use the \bagp software \citep{Carnall:2018gb} to fit the BC03 \citep{Bruzual:2003ck} models of SEDs to the photometric measurements derived above. 
We assume the \citet{Chabrier:2003ki} initial mass function, 
a metallicity range of $Z/Z_{\odot}\in(0, 2.5)$, 
the \citet{Calzetti:2000iy} extinction law with $A_v$ in the range of (0, 3). \,
We use the Double Power Law (DPL) model other than simple exponentially declining form to capture the complex Star Formation History (SFH) of our galaxies at cosmic noon (rather than local universe), following \cite{Carnall:2019di}.
The nebular emission component is also added into the SED during the fit, since our galaxies are exclusively strong line emitters by selection.
The redshifts of our galaxies are fixed to their best-fit grism values, with a conservative uncertainty of $z_{\sigma} = 0.003$.
Note that we have obtained the entire redshift posterior from \grzl in Sec:\ref{subsec:gri}, and set a criterion of $\left(\Delta z\right)_{\rm posterior}/(1+z_{\rm peak})<0.002$ for secure redshift measurements.
But here we still set a Gaussian prior centered on $z_\text{peak}$ with $z_{\sigma} = 0.003$ for simplicity in SED fitting, following \citet{Momcheva:2016fr}.
Actually, the minimum, median, and maximum values of $\Delta z/(1+z)$ for our sample are $1.4\times10^{-4}, 2.8\times10^{-4}, 1.5\times10^{-3}$, respectively.

Our mass estimates are in agreement with \citet{Santini:2023ApJ...942L..27S}, even though we stress that our results are more robust, because we use spectroscopic redshifts.
After correcting magnification according to our recent lensing model \citep{Bergamini:2023arXiv230310210B}, we are allowed to take a glimpse of the loci of our galaxies in the SFR–M* diagram as in Fig.~\ref{fig:sfr}.
We show the star-forming main sequence fitted by \citet{Speagle:2014dd}, which is extrapolated from $\log (M_*/M_\odot)\in[9.7,11.1]$ to the mass range of our sample with $\pm0.2$ dex scatters. 
\citet{Sanders:2021ga} gives stacked results of field galaxies fairly close to their extrapolated best fit out to $\log(M_*/M_\odot)=9$.
Our sample generally scatters around the main sequence at higher $M_*$.
But at lower $M_*$ high SFR galaxies are dominant, especially for $z\sim 3$ at $M_*/M_\odot\lesssim3\times10^8$.
It might account for the low metallicity at the low mass region when assuming the FMR \citep{Mannucci:2010MNRAS.408.2115M},  which will be discussed in Sect.~\ref{subsec:mzr}.

\subsection{AGN contamination}
\label{subsec:AGN}

\begin{figure}
    \centering
    \includegraphics[width=\columnwidth]{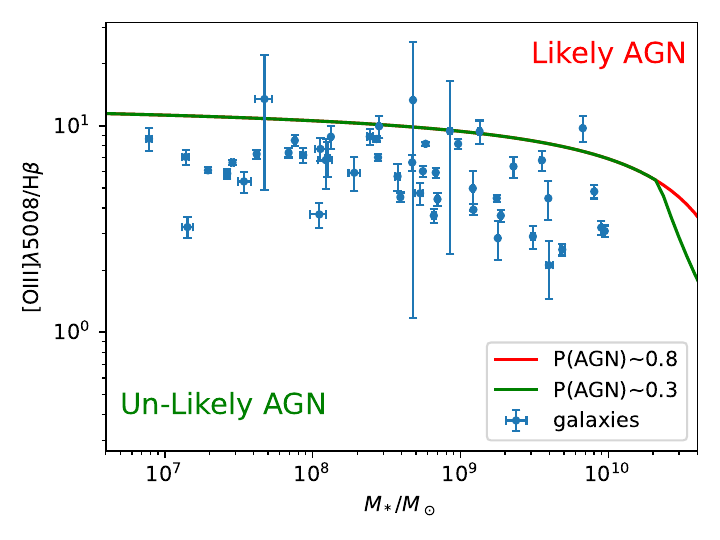}
    \caption{The Mass-Excitation diagram of our sample, used to exclude possible AGN galaxies. 
    The position of the likely AGN galaxies with the possibility of 0.8 and 0.3, are marked by the red and green curves.
    No significant possible AGN contamination is evident in our samples, with one galaxy (ID=03854) only slightly off by $1\sigma$.}
    \label{fig:MEx-AGN}
\end{figure}

The metallicity diagnostics used in this work are strictly for star-forming regions/galaxies, and the results will be incorrect if there is Active Galactic Nucleus (AGN) emission.
So the last step is to exclude the AGN contamination from purely star-forming galaxies, by using the mass–excitation (MEx) diagram as shown in Fig.~\ref{fig:MEx-AGN}.

AGNs leave strong signatures on nebular line ratios such as $\OIII~\lambda5007/\Hb$ and/or $\NII~\lambda6584/\Ha$, which form the most traditional version of the BPT diagram \citep{Baldwin:1981ev}.
Due to the limited spectral resolution of JWST/NIRISS slitless spectroscopy ($R\sim150$), \NII is entirely blended with \Ha, which precludes us from using the BPT diagram to remove AGN contamination.

Fortunately, \citet{Juneau:2014ca} proposed an effective approach coined the mass-excitation (MEx) diagram, using \Mstar as a proxy for \NII/\Ha, which functions well at $z\sim0$ (\ie SDSS DR7).
\citet{Coil:2015dp} further modified the MEx demarcation by horizontally shifting these curves to high-\Mstar by 0.75 dex, which is shown to be more applicable to the MOSDEF sample \citep{Sanders:2021ga} at $z\sim2.3$. %  (\ie the MOSDEF sample).
We thus rely on this modified MEx to prune AGN contamination from our galaxy sample.
As shown in Fig.~\ref{fig:MEx-AGN}, the green and red curves mark the steep gradient of $P(AGN)\sim0.3$ and $P(AGN)\sim0.8$ respectively, which represent the probability that the galaxy hosts an AGN. 

Most of the sources are clearly un-likely AGN, and some scattered around the critical line are ambiguous.
There are only two galaxies slightly above the upper demarcation within $1\sigma$.
Because our analysis is based on stacking, a small minority of contaminating AGN will have a negligible impact.
Given the limited sample size, we tend to retain more applicable data, and consequently, no possible AGN is eliminated and we preserve all 50 galaxies.

\subsection{Stacking spectra} \label{subsec:stack}

%= = = = = = = = = = = = = = = = = = = = = = = = = = = = = = = = = = = = = = = =
% = = = = = = = = = = = = = = = = = = = = = = = = = = = = = = = = = = = = = = = = = =
% Include this table with \input{filename.tex}
% To rotate in emulateapj do: \begin{turnpage}\input{filename.tex}\end{turnpage}
% To display it on multiple pages do: \LongTables\input{filename.tex}
% - - - - - - - - - - - - - - - - - - - - - - - - - - - - - - - - - - - - - - - - - -
{
\tabletypesize{\scriptsize}
\tabcolsep=2pt
\begin{deluxetable*}{lccccccccccc}
    \tablecolumns{12}
    \tablewidth{0pt}
    \tablecaption{Measured properties of the stacked spectra.}
% - - - - - - - - - - - - - - - - - - - - - - - - - - - - - - - - - - - - - - - - - -
\tablehead{
    \colhead{group} &
    %\colhead{z range}   &    
    \colhead{$N_\text{gal}$} &
    \colhead{mass range} &
    \colhead{log$M_{\ast}^{\rm med}$} & % /M_{\odot} \tablenotemark{a} &
    \colhead{\OIII/\Hb}  &
    \colhead{\OII/\Hb}  &
    \colhead{\OIII/\OII}  &
    \colhead{\Hg/\Hb}   &
    % \colhead{$f_{\OIII}$}  &
    \colhead{\NeIII/\OIII}   &
    \colhead{\Ha/\Hb}   &
    \colhead{\SII/\Ha}  &
    \colhead{\oh} % \tablenotemark{b} \\
%\colhead{$\Delta~\mathrm{log(O/H)_{cluster-field}}$\tablenotemark{c}}  &
    % \colhead{\oh\tablenotemark{d}}  \\
    % & range & [$M_{\odot}$] & [$M_{\odot}$/yr] &  &  &  &  & [$10^{-17}$\Funit]  &  \multicolumn{2}{c}{using B18 calibrations}  &  using C17 calibrations
}
%---------------------------------------------------------------
\startdata
    \multicolumn{12}{c}{$1.8<z_\mathrm{grism}<2.3$} \\
11 & 7 & [6.8,7.7) & 7.42 & $7.21 \pm 0.55$ & $0.75 \pm 0.07$ & $9.62 \pm 0.58$ & $0.26 \pm 0.04$ & $0.07 \pm 0.01$ & $2.88 \pm 0.26$ & $0.01 \pm 0.02$ & $8.00_{-0.04}^{+0.05}$ \\
12 & 10 & [7.7,8.7) & 8.20 & $7.27 \pm 0.63$ & $1.40 \pm 0.16$ & $5.20 \pm 0.44$ & $0.13 \pm 0.07$ & $0.01 \pm 0.02$ & $2.77 \pm 0.29$ & $0.08 \pm 0.03$ & $8.15_{-0.05}^{+0.05}$ \\
13 & 11 & [8.7,9.9) & 9.09 & $4.84 \pm 0.23$ & $1.93 \pm 0.15$ & $2.51 \pm 0.17$ & $0.09 \pm 0.05$ & $0.02 \pm 0.02$ & $3.56 \pm 0.24$ & $0.09 \pm 0.03$ & $8.37_{-0.05}^{+0.04}$ \\
\hline
\multicolumn{12}{c}{$2.6<z_\mathrm{grism}<3.4$}\\
21 & 5 & [7.1,8.2) & 7.84 & $5.10 \pm 0.64$ & $0.32 \pm 0.07$ & $15.83 \pm 2.94$ & ... & $0.04 \pm 0.01$ & ... & ... & $7.98_{-0.12}^{+0.19}$ \\
22 & 9 & [8.2,9.2) & 8.84 & $7.48 \pm 0.55$ & $1.75 \pm 0.15$ & $4.29 \pm 0.23$ & $0.23 \pm 0.15$ & $0.03 \pm 0.01$ & ... & ... & $8.25_{-0.06}^{+0.05}$ \\
23 & 8 & [9.2,10.0) & 9.57 & $3.91 \pm 0.33$ & $1.80 \pm 0.24$ & $2.17 \pm 0.22$ & $0.28 \pm 0.16$ & $0.05 \pm 0.04$ & ... & ... & $8.47_{-0.06}^{+0.05}$ \\
\enddata
% - - - - - - - - - - - - - - - - - - - - - - - - - - - - - - - - - - - - - - - - - -
    \tablecomments{The multiple emission line flux ratios are measured from the stacked spectra shown in Fig.\ref{fig:stack}. 
    The mass range and the median stellar mass $\log M_*^\mathrm{med}$ are both logarithmic values $\log (M_*/M_\odot)$.
    The metallicity inference is derived from the measured line flux ratios in the stacked spectra presented in each corresponding row, using the method described in Sect.~\ref{subsec:metal}.
    Here we use the strong line calibrations prescribed by \citet[][B18]{Bian:2018km} and some others. % as our default results. 
    See Table~\ref{tab:coef} for the relevant coefficients.
    % The $f_{\OIII}$ and SFR results refer to the median value of galaxies within each mass bin, with 1-$\sigma$ uncertainty represented by the standard deviation.
    }
    %\tablenotetext{a}{The median stellar mass of galaxies within each mass bin.}
%    \tablenotetext{b}{The metallicity inference derived from the measured line flux ratios in the stacked spectra presented in each corresponding row, using the method described in Sect.~\ref{subsec:metal}.
    %Here we use the strong line calibrations prescribed by \citet[][B18]{Bian:2018km} and some others. % as our default results. 
   % See Table~\ref{tab:coef} for the relevant coefficients.}
    %\tablenotetext{c}{The difference in metallicity between our galaxies in overdense environments and the field measurements inferred from the fundamental metallicity relation prescribed by % \citet{Sanders:2021ga}.
    %The intrinsic scatter of 0.06 dex has been combined in quadrature into the measurement uncertainties.}
    %\tablenotetext{d}{The metallicity inference derived using the strong line calibrations given by %\citet{Curti:2017fn}. All other assumptions and data are the same as our default results using the B18 calibrations.}
\label{tab:stack}
\end{deluxetable*}
}
%= = = = = = = = = = = = = = = = = = = = = = = = = = = = = = = = = = = = = = = =

\begin{figure*}
    \centering
    \includegraphics[width=\textwidth]{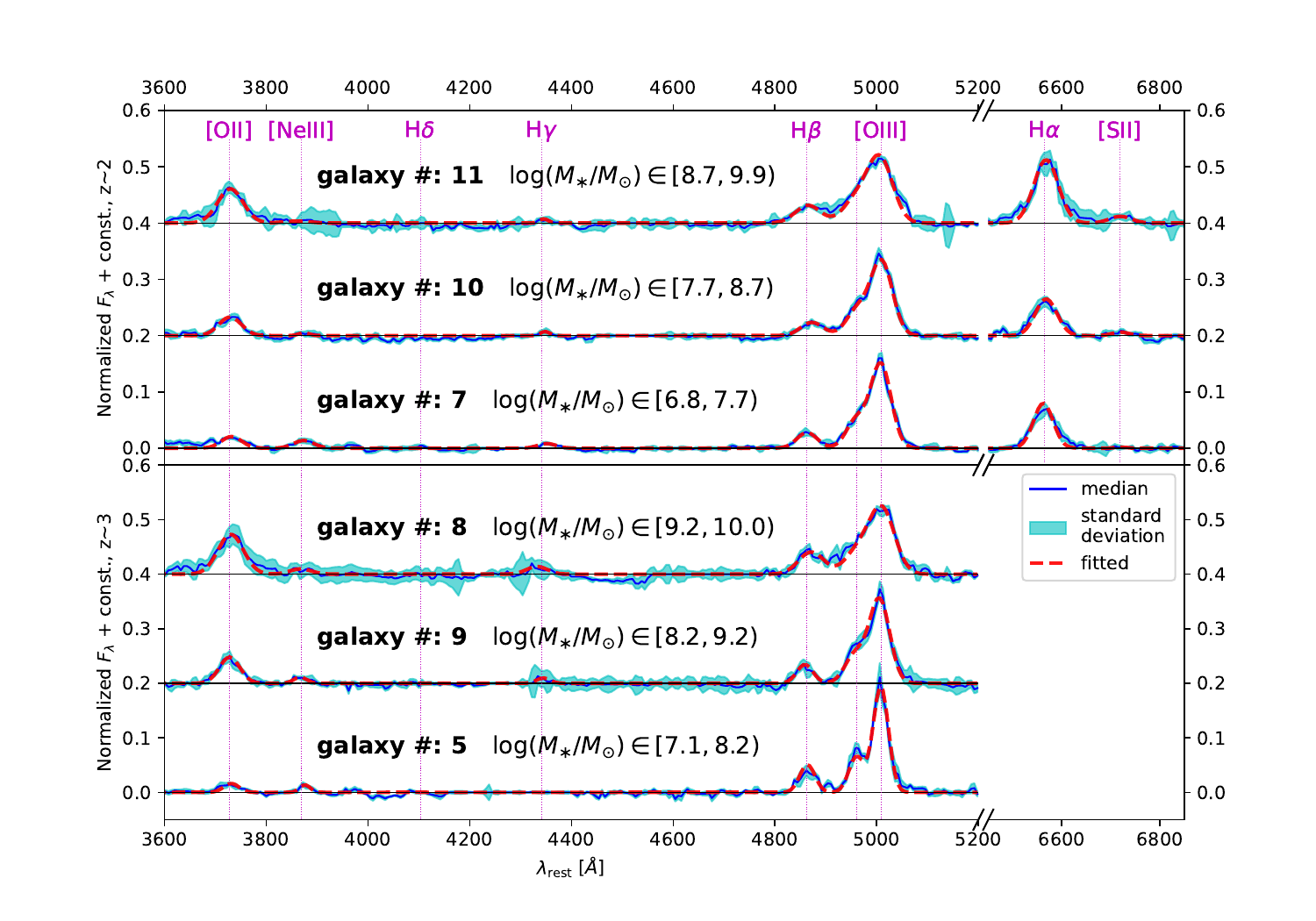}
    \caption{Stacked grism spectra for galaxies residing in several mass bins at two redshift ranges, as shown in the upper $(1.8<z<2.3)$ and lower $(2.6<z<3.4)$ panels, respectively.
    Each mass bin contains $5\sim 11$ galaxies, with the exact number of galaxies and corresponding mass range highlighted above each stacked spectrum.
    In each set of spectra, the blue curves represent the median stacked spectrum, the cyan bands mark the standard deviation % bootstrapped 
    flux uncertainties, and the red dashed curves show the best-fit Gaussian fits to multiple emission lines, while \SII, \Ha are across a discontinuous range among other lines (\ie, the \OIII$\lambda\lambda$4960,5008 doublets, \Hb, \Hg, \Hd, \NeIII, and \OII) in the broken axes at right parts.
    The details of the stacking procedures are presented in Sect.~\ref{subsec:stack}.
    }
    \label{fig:stack}
\end{figure*}

Robust emission lines are required to estimate metallicity for MZR measurement.
So we need composite spectra obtained by stacking procedure to achieve higher SNR from low-resolution grism spectra.
In the previous subsection, we have selected 50 spectroscopically confirmed galaxies in the A2744 lensed field that are undergoing active star formation.
Then they are divided into 2 redshift bins ($z\in[1.8, 2.3]$ and $z\in[2.6, 3.4]$), and 3 mass bins respectively as in Tab.~\ref{tab:stack}.
Our choice of binning aims to have a reasonable number of galaxies per bin.
We tested that changing the mass bins does not significantly affect our conclusions.
Approximately each mass bin contains $\sim7$ individual galaxies, and the SNR will be increased roughly by a factor of $\sqrt{7}=2.6$.
The 1D/2D spectra of representative galaxies in each of the 6 bins are shown in Fig.~\ref{fig:indivi_spec_example}.

Then we adopt the following stacking procedures, similar to those utilized by \citet{Henry:2021ju,Wang:2022ApJ...926...70W}:

\begin{enumerate}
    \item Subtract continuum models from the extracted grism spectra. The continua are constructed by \textsc{Grizli} combining two orients. We apply a multiplicative factor to the continuum models to make sure there is no offset between the modeled and observed continuum levels around emission lines, to avoid continuum over-subtraction.
    \item Normalize the continuum-subtracted spectrum of each object using its measured \OIII flux, to avoid excessive weighting toward objects with stronger line fluxes. Here the \OIII fluxes we used are the results of 1D line profile fitting instead of 2D forward modeling by \textsc{Grizli}, for a more straightforward normalization.
    \item De-redshift each normalized spectrum to its rest frame, and resample on the same wavelength grid using \textsc{SpectRes}\footnote{\url{https://spectres.readthedocs.io/en/latest/}} with the integrated flux preservation.
    \item Take the median and the variance of the normalized fluxes at each wavelength grid, as the value and uncertainty of the stacked spectrum.
\end{enumerate}

As shown in Fig.~\ref{fig:stack}, these key lines are more significant in stacked spectra.
The (relative) emission line fluxes are measured by fitting a set of Gaussian profiles to the line in stacked spectra, as well as individual spectra.
We simultaneously fit \OII, \NeIII, \Hd, \Hg, \Hb, \OIII, \Ha, and \SII.
The amplitude ratio of $\OIII~\lambda\lambda4960,5008$ doublets is fixed to 1:2.98 following \citet{Storey:2000jd}.
The centroids of Gaussian profiles are allowed a small shift of the corresponding rest-frame
wavelengths of emission lines, within ±10 \AA, in order to accommodate systematic uncertainties.
The FWHMs of each line are not required to be the same, but set between $[10,25]\AA$, consistent with the rest-frame spectral resolution $\Delta\lambda\approx 7 \AA$ corresponding to $R\approx150$ for NIRISS.  
We use the software \textsc{LMFit}\footnote{https://lmfit.github.io/lmfit-py/} to perform the nonlinear least-squares
minimization, with the measured quantities summarized in Tab.~\ref{tab:stack}.
The stacked metallicity is estimated using the same methods as the individual galaxies outlined in Sect.~\ref{subsec:metal}.
Our later discussion will mainly focus on the stacked results. 

\section{Results}\label{sec:result}

From the joint analysis of the JWST/NIRISS and JWST/NIRCam data, we revisit the measurement of the MZR using the stacked spectra of the A2744-lensed field galaxies within the mass range of $M_*\in(10^{6.9},10^{10.0})M_{\odot}$ at $z\in(1.8,3.4)$, shown in Sect.\ref{subsec:mzr}.
We also perform a systematic investigation of the differences between 2D and 1D forward modeled fluxes of nebular emission lines from slitless spectroscopy, as detailed in Sect.~\ref{subsec:lineflux}.

% = = = = = = = = = = = = = = = = = = = = = = = = = = = = = = = = = = = = = = = = = =
% Include this table with \input{filename.tex}
% To rotate in emulateapj do: \begin{turnpage}\input{filename.tex}\end{turnpage}
% To display it on multiple pages do: \LongTables\input{filename.tex}
% - - - - - - - - - - - - - - - - - - - - - - - - - - - - - - - - - - - - - - - - - -
%{
%\tabletypesize{\scriptsize}
%\tabcolsep=2pt
\begin{deluxetable}{lcllccccccccc}
    \tablecolumns{5}
    \tablewidth{0pt}
    \tablecaption{Comparison of MZR from different works, which is defined as: $12 + \log(\mathrm{O/H}) = \beta \times\log(M_*/10^8M_\odot)+ Z_8$.}
% - - - - - - - - - - - - - - - - - - - - - - - - - - - - - - - - - - - - - - - - - -
\tablehead{
    \colhead{Papers} &
    \colhead{$z_\text{median}$} &
    \colhead{slope $\beta$} &
    \colhead{intercept $Z_8$}& %\tablenotemark{b} &
    \colhead{calibration}
}
%---------------------------------------------------------------
\startdata
    %\multicolumn{3}{c}{Strong line calibrations of \citet[][B18]{Bian:2018km}\tablenotemark{a}}   \\
    \noalign{\smallskip}
    \multirow{2}{*}{this work, stack} & 1.90 & $0.223 \pm 0.017$ & $8.123 \pm 0.012$ &  \\
    % \multirow{2}{*}{B18} \\
    & 2.88 & $0.294 \pm 0.010$ & $8.008 \pm 0.013$ & \citet{Bian:2018km} \\
    \multirow{2}{*}{individual} & 1.90 & $0.229 \pm 0.028$ & $8.079 \pm 0.027$ & hereafter, B18\\
    & 2.88 & $0.295 \pm 0.043$ & $7.981 \pm 0.051$ & \\
    \noalign{\smallskip}\hline\noalign{\smallskip}
    \multirow{2}{*}{this work, stack} & 1.90 & $0.314 \pm 0.053$ & $8.064 \pm 0.043$ & \multirow{2}{*}{\citet{Sanders:2023arXiv230308149S}} \\
    & 2.88 & $0.586 \pm 0.051$ & $7.748 \pm 0.059$ & \\
    \noalign{\smallskip}\hline\noalign{\smallskip}
    \multirow{2}{*}{\citet{Li:2022arXiv221101382L}} & 2 & $0.16\pm0.02$ & $8.18\pm0.03$ & B18 \\
    % \multirow{2}{*}{B18, $\rm{O}_{32}$} \\
     & 3 & $0.16\pm0.01$ & $8.08\pm0.01$ & only $\rm{O}_{32}$ \\
    \noalign{\smallskip}\hline\noalign{\smallskip}
    \multirow{2}{*}{\citet{Sanders:2021ga}} & 2.2 & $0.30\pm0.02$ & $7.91\pm0.04$ & \multirow{2}{*}{B18} \\
     & 3.3 & $0.29\pm0.02$ & $7.83\pm0.04$ & \\
    \noalign{\smallskip}\hline\noalign{\smallskip}
    \citet{Henry:2021ju} & 1.9 & $0.22\pm0.03$ & $7.98\pm0.06^*$ & \citet{Curti:2017fn} \\
    \noalign{\smallskip}\hline\noalign{\smallskip}
    \citet{Wang:2022ApJ...926...70W} & 2.2 &  $0.14\pm0.02$ & $8.17\pm0.03$ & B18 \\
    \noalign{\smallskip}\hline\noalign{\smallskip}
    \citet{Heintz:2023NatAs} & 7-10 &  $0.33$ & $7.29$ & \citet{Sanders:2023arXiv230308149S} \\
    \noalign{\smallskip}\hline\noalign{\smallskip}
    \citet{Curti:2023arXiv230408516C} & 3-6 & $0.21\pm0.04$ & $7.80\pm0.03$ &  \citet{Laseter:2023arXiv} %\cite{Curti:2020MNRAS.491..944C}
\\
    \noalign{\smallskip}\hline\noalign{\smallskip}
    \citet{Nakajima:2023arXiv230112825N} & 4-10 & $0.25\pm0.03$ & $7.74\pm0.06^*$ & \citet{Nakajima:2022ApJS..262....3N} \\
    \noalign{\smallskip}\hline\noalign{\smallskip}
    %\multicolumn{5}{c}{Strong line calibrations of \citet[][C17]{Curti:2017fn}\tablenotemark{b}}   \\
    %\noalign{\smallskip}
    %\OIII/\Hb   &   -0.277 &  -3.549 & -3.593 & -0.981  \\
    %\OII/\Hb    &   0.418 &  -0.961 & -3.505 & -1.949   \\
    %\multicolumn{5}{c}{Balmer decrement}   \\
    %\noalign{\smallskip}
    %\Hg/\Hb   &   -0.3279    &   \nodata   &   \nodata  &   \nodata \\
\enddata
% - - - - - - - - - - - - - - - - - - - - - - - - - - - - - - - - - - - - - - - - - -
    % \tablecomments{The empirical flux ratios are computed in the polynomial functional form of $\log{R} = \sum_{i} c_i\cdot x^i$, where $x=\oh-8.69$ for the \citet{Curti:2017fn} calibrations and $x=\oh$ for the \citet{Bian:2018km} ones. }
    %\tablenotetext{b}
    \tablecomments{The intercept provided in \citet{Sanders:2021ga} \& \citet{Nakajima:2023arXiv230112825N} is $\text{Z}_{10}$ instead of $\text{Z}_{8}$, where $Z_8=-2\beta+Z_{10}$. The errors they correspond to (marked by $*$) are only conservative upper limits: $\sigma_8=\sqrt{(-2)^2\sigma_{\beta}^2+\sigma_{10}^2+2\sigma_{\beta Z_{10}}}$, since we do not know the (negative) covariance $\sigma_{\beta Z_{10}}$ therein.}
    %{The intrinsic scatter of 0.09(0.11) dex for the \OIII/\Hb(\OII/\Hb) calibrations, as reported in \citet{Curti:2017fn}, have been included in the metallicity inference.}
\label{tab:mzr}
\end{deluxetable}
%}

\begin{figure*}
    \centering
    \includegraphics[width=\textwidth]{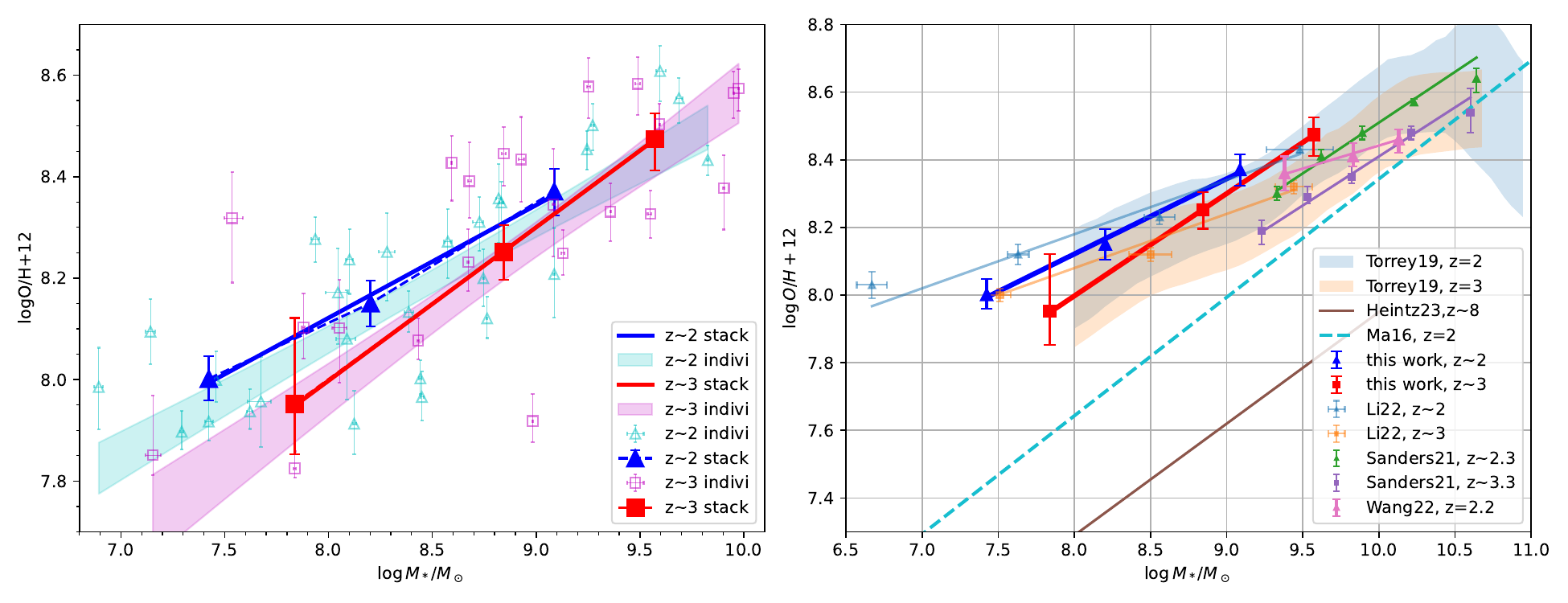}
    \caption{
    MZR measurements for the star-forming field galaxies behind the A2744 cluster.
    \textit{Left:} the individual (hollow) and the stacked (solid) result of our galaxy sample at $z\in[1.8,2.3]$ (blue triangles) and $z\in[2.6,3.4]$ (red squares), with their linear fits represented by shaded regions and solid lines.
    \textit{Right:} comparison to other observational works, along with the IllustrisTNG100 simulation \citep{Torrey:2019MNRAS.484.5587T} and the FIRE simulation \citep{Ma:2016gw}.
    These colored lines are linear regressions of their respective results, with their parameters summarized in Tab.~\ref{tab:mzr}.
    }
    \label{fig:mzr}
\end{figure*}

\subsection{The MZR at the low mass end}\label{subsec:mzr}

Our key scientific result is the measurement of the gas-phase MZR in the low mass range of $\log (M_*/M_{\odot})\in(6.9, 10.0)$ at $z\in(1.8,3.4)$.
The slope of the MZR has been shown to be a key diagnostic of galaxy chemical evolution and the cycling of baryons and metals through star formation and gas flows \citep[see \eg,][and references therein]{Maiolino:2019vq}.
In particular, \citet{Sanders:2021ga} argues that the shape of the MZR at $z\sim2-3$ is more tightly regulated by the efficiency of metal removal by gas outflows $\zeta_\text{out}$, rather than by the change of gas fractions with stellar mass $\mu_\text{gas}(M_*)$. \citet{Henry:2013cn} observes a steepening of the MZR slope at $z\sim2$, suggesting a transition from momentum-driven winds to energy-driven winds as the primary prescription for galactic outflows in the low-mass end.

We find a clear correlation between metallicity and stellar mass for both individual galaxies and stacked spectra at $z\in[1.8,2.3]$ and $z\in[2.6,3.4]$, as shown in the left panel of Fig.\ref{fig:mzr}.
The $z\sim2$ and $z\sim3$ individual galaxy samples have Spearman correlation coefficients of 0.788, 0.688 with the $p-\text{value}$ of $6.36\times10^{-7},3.98\times10^{-4}$, respectively.
We perform linear regression over the stacks to derive the MZR:
\begin{equation}
    12 + \log(\mathrm{O/H}) = \beta \times\log(M_*/10^8M_\odot)+ Z_8,
\end{equation}
where $\beta$ is the slope and $Z_8$ is the normalization at $M_* = 10^8 M_\odot$, as the blue and red solid line with uncertainties at $z\sim2,3$ in both panels of Fig.\ref{fig:mzr}.
We measure the MZR slope to be $\beta=0.223 \pm 0.017$ and $\beta=0.294 \pm 0.010$ for our galaxy samples at $z_{\rm median}=1.90$ and $z_{\rm median}=2.88$, respectively.
We see moderate evolution in the MZR normalization from $z\sim2$ to $z\sim3$: $\Delta Z_8=-0.11\pm0.02$.
The stacked MZRs demonstrate good agreement with the individual results (linear fits are shown in the shaded regions in the left panel of Fig.~\ref{fig:mzr}.)
The large uncertainty of the stacked metallicity in the $z\sim3$ lowest mass bin, comes from the limited number of galaxies.
More importantly, all 5 galaxies within this bin are high-SFR galaxies (Fig.~\ref{fig:sfr}), which might explain their low stacked metallicity, under the assumption that the star-forming main sequence \citep{Speagle:2014dd} and the FMR \citep{Mannucci:2010MNRAS.408.2115M} are valid below $M_*\lesssim8$.
A detailed study and characterization of incompleteness at the low mass end is beyond the scope of this paper, and is left for future work.

We summarize our measurements in Table~\ref{tab:mzr}, along with other literature results.
The right panel of Fig.\ref{fig:mzr} shows the comparisons to other observations and two cosmological hydrodynamic simulations.
In addition to $z\sim2,3$, we also include 3 latest MZR measurements at a very high redshift from JWST/NIRSpec for comparison. 
We measure the slope of the MZR to be $\beta\sim0.25$ for both $z\sim2$ and $z\sim3$.
Our slopes at low mass are slightly lower than those found by \citet{Sanders:2021ga}, but ours are in lower mass ranges.
The shallower normalization could be accounted for the MZR evolution from ours $z_\text{median}=1.90,2.88$ to theirs $z\sim2.3,3.3$.
Furthermore, we follow their analytic model to understand what physical processes set the slope at the dwarf mass range. 
In the \citet{Peeples:2011ew} model, the metallicity of the ISM is expressed as:
\begin{equation}
    Z_\mathrm{ISM} = \frac{y}{\zeta_\mathrm{out}-\zeta_\mathrm{in}+\alpha\mu_\mathrm{gas}+1}.
\end{equation}
Following the assumption by \citet{Sanders:2021ga} that the gas fraction $\mu_\mathrm{gas}=10^{\mu_0}{M_*}^{-0.36}$ ($\mu_0=3.89,3.96$ for $z\sim2,3$, respectively), the coefficient $\alpha=0.7\cdot(0.64+\beta)$, the nucleosynthetic stellar yield $ y/Z_\mathrm{ISM}=10^{9.2-(12+\log(\mathrm{O/H}))}$ the metal loading factors of inflowing gas accretion $\zeta_\mathrm{in}=0$, we calculate the loading factors of outflowing galactic winds $\zeta_\mathrm{out}$ at each stacked point and linear fit. We get:
\begin{equation}
    \begin{aligned}
    z\sim2: \log(\zeta_\mathrm{out}) =& (-0.130\pm0.072) m_{10} + (0.408\pm0.119),  \\
    z\sim3: \log(\zeta_\mathrm{out}) =& (-0.332\pm0.037) m_{10} +(0.202\pm0.035),  \\
    \text{ where, }\quad m_{10}=&\log(M_*/10^{10}M_\odot).
    \end{aligned}
\end{equation}
And we find that $\log(\zeta_\mathrm{out}/\alpha\mu_\mathrm{gas})$ is only a little bit above zero over the mass range, with $\zeta_\mathrm{out} \approx 1.01-1.5\times \alpha\mu_\mathrm{gas}$.
Thus our results indicate that the shallower MZR may be attributed to a shallower $M_*$ scaling of the metal loading of the galactic outflows $\zeta_\text{out}$ at the low mass end.
We generalize their conclusions that outflows $\zeta_\text{out}$ remain the dominant mechanism other than gas fraction $\mu_\text{gas}$ that sets the MZR slope, and $\mu_\text{gas}$ gradually carries more relative importance and rise to nearly the same order as $\zeta_\text{out}$ for the low mass regime.

Our MZR slope $\beta\sim0.25$ is steeper than those reported in \citet{Li:2022arXiv221101382L} at the same redshifts and similar mass range as in Tab.~\ref{tab:mzr}.
Although we use the same NIRISS data of the A2744 lensed field, we only match 28 out of 50 galaxies with the on-sky distances(d2d) lower than 1 arcsec to the Abell catalogue of \citet{Li:2022arXiv221101382L}, and only 18/50 of them are in agreement with our metallicity measurements within $1\sigma$ confidence interval.
This difference likely arises from the updated calibration files used in our NIRISS data reduction, and from our Bayesian approach in the metallicity inference using multiple line ratios to joint fit other than only \OIII/\OII from \citet{Bian:2018km}.
In addition, we include the new JWST/NIRCam imaging data covering the rest-frame optical wavelength ranges for our sample galaxies \citep{Paris:2023arXiv230102179P}, use more complex SFH (DPL), and employ the latest JWST-based lensing model \citep{Bergamini:2023arXiv230310210B} for more reliable stellar mass estimates.
Another source of difference is their choice of exponentially declining SFH ($\tau$ model) which may not be appropriate for our high-redshift star-forming galaxies \citep{Reddy:2012hw}, and might introduce a significant bias in stellar mass $M_*$ estimation \citep{Pacifici:2015gg,Carnall:2018gb,Carnall:2019di}.

In agreement with previous work, we also find a tendency for the slope of the MZR to flatten out in the low mass at around $M_*/M_\odot\lesssim10^9$, although not as significant.
As for higher redshift $z\sim3-10$, our inferred slopes $\beta$ are consistent with those by \citet{Curti:2023arXiv230408516C, Nakajima:2023arXiv230112825N}, but our intercept $Z_8$ are $\sim0.3$ dex higher.
At that time, the metal might be enriching and hence the MZR might be building up \citep{Curti:2023arXiv230408516C}, and it is not until the SFR peaks at "cosmic noon" $z\sim2-3$ that the MZR exhibits a higher intercept.

The MZR measurements are also sensitive to different strong line calibrations, especially for the intercept $Z_8$ \citep{Kewley:2008be}, as discussed in Sect.\ref{subsec:metal}.
In Tab.~\ref{tab:mzr}, we also provide the MZR from stacks using the \citet{Sanders:2023arXiv230308149S} calibration for comparison.
Although the measured slopes are significantly steeper than our default B18 MZR, they are still consistent with \citet{Heintz:2023NatAs} for dwarf galaxies at higher redshift.
We fit the stacked result presented by \citet{Henry:2021ju} in the similar mass range, which assumes \citet{Curti:2017fn} calibration.
Our slope agrees with theirs $\beta=0.22\pm0.03$, but the intercept is $\sim0.1$ dex higher.
This agrees with \citet{Wang:2022ApJ...926...70W,Li:2022arXiv221101382L}, who test that the calibrations of \citet{Bian:2018km} yielded a steeper MZR than the calibrations of \citet{Curti:2017fn}  when analyzing the same data.

Moreover, we compare our results with two simulation works presented separately in Fig.~\ref{fig:mzr}.
Our individual measurements are largely compatible with the result of the simulations \textsc{IllustrisTNG} \citep{Torrey:2019MNRAS.484.5587T}.
But several high metallicity galaxies lift the stacked MZR up high slightly, yielding a steeper slope than they predicted.
Our measured slopes are in better agreement with the \textsc{FIRE} simulation results \citep{Ma:2016gw}, which are capable of resolving high-$z$ dwarf galaxies with sufficient spatial resolution. 

In addition, all the MZRs discussed above are derived from galaxy populations residing in random fields. There has been continuous discussion about the environmental dependence of MZR shapes at high redshifts \citep{Peng:2014MNRAS, Bahe:2017MNRAS, Calabro:2022A&A,Wang.2023}.
Here we raise one recent observation of the MZR at $z\sim2.2$ showing a much shallower slope ($\beta=0.14\pm0.02$), measured using the HST grism spectroscopy of 36 galaxies residing in the core of the massive BOSS1244 protocluster \citep{Wang:2022ApJ...926...70W}. 
Our work presented here confirms the significant difference between the MZR slopes measured in field and overdense environments, indicating the change in metal removal efficiency as a function of the environment.

\subsection{Investigation of the morphological broadening effect on measurements of line flux and metallicity}\label{subsec:lineflux}

\begin{figure*}
    \centering
    \includegraphics[width=\textwidth]{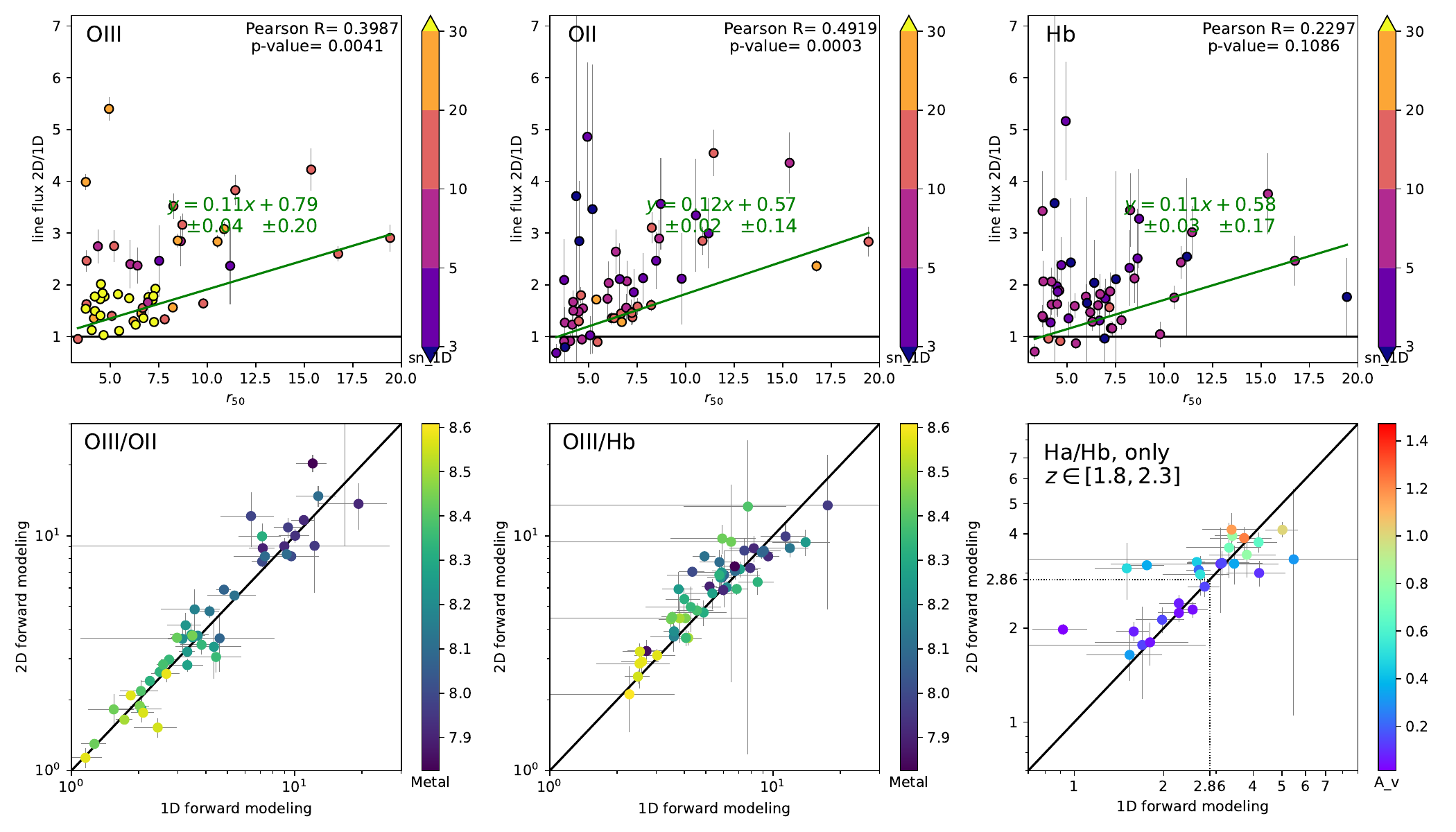}
    \caption{
    Comparison between the emission line fluxes derived using the 2D/1D forward modeling methods, explained in detail in Sect.\ref{subsec:gri} \& \ref{subsec:stack}, respectively.
    The top 3 panels show the galaxy radius vs. the flux ratio of 2D to 1D for each line.
    The 2D fluxes are tangibly higher than 1D fluxes (above the black line), and it seems systematic for all 3 brightest lines of each source (with the same $r_{50}$).
    We find a correlation between them (green line), although not so strong, with the Pearson correlation coefficients and the p-value exhibited in the top right corner, as well as the green result of linear fitting at the center. 
    Their color marks the SNR of the flux from the 1D method, showing no significant correlation.
    The bottom 3 panels show the line flux ratio, while their color marks the metallicity or the dust extinction derived in Sect.\ref{subsec:metal} using \grzl flux ratio.
    These two distributions nearly scatter across the equality line (in black) within the uncertainty. But there are several outliers and a slight systematic overestimation for 2D, which is more obvious for \Ha/\Hb at the bottom right.
    }
    \label{fig:compareflux}
\end{figure*}

Since metallicity estimates heavily rely on line flux measurements, in this section we verify that different methodologies in deriving emission line fluxes from the NIRISS slitless spectroscopy with limited spectral resolution do not result in significant biases on the metallicity derivations. 

For grism spectroscopy, it has long been recognized that the morphological broadening effect can change the overall spectral shape and flux levels of galaxies
\citep[see \eg, ][]{vanDokkum:2011cq,Wang:2019cf,Wang:2020bp}.
We thus systematically compare, for the first time, two methods to measure emission line flux from slitless spectroscopy, with and without the consideration of this morphological broadening effect. The 2D forward modeling analysis of \grzl is depicted in Sect.\ref{subsec:gri}. In this section, we describe the line profile fitting to 1D extracted spectra using \lmfit. The morphology of a galaxy has already been taken into account when forward modeling its 2D spectrum by \grzl.
% and the line flux derived by \grzl are intrinsic.
The extracted 1D spectra are morphologically broadened along the dispersion direction, and can vary significantly in spectral slope and flux level for the same object due to the different projected 1D morphology \citep[see Fig. 8,9 of][for examples]{Wang:2019cf}.
Therefore, we regard the 2D line flux as the reference intrinsic value, and 1D flux as the measurement not corrected for the morphology.
The difference has not yet been fully investigated, and thus demands immediate attention, with the upcoming advent of large slitless spectroscopic surveys, \eg, Euclid, Roman, and the Chinese Space Station Telescope (CSST).

In the top 3 panels of Fig.\ref{fig:compareflux}, we show the comparison between line flux measured from 2D or 1D spectra, and try to associate it with the half-light radius $r_{50}$.
The flux ratio of 2D to 1D deviates from 1 tangibly, and 2D flux modeled by \grzl are larger in most cases (47/48, 41/48, 43/48 for \OIII, \OII, \Hb, respectively) than 1D flux fitted using \lmfit by a median factor of $\sim30\%$ (with wide dispersion -0.3 -- 5, where minus factor means 2D flux is lower than 1D flux).
This strong offset does not seem to be related to SNR.
As expected, we find it does correlate with the half-light radius $r_{50}$ of the individual galaxies, although not as strong as the Pearson correlation coefficients $R$ shown.
The unit of $r_{50}$ is the pixel, and here 1 pixel corresponds to 0.03 arcsec, as illustrated in Sect.~\ref{sec:data}.
Furthermore, Pearson $R$ decreases as the SNR decreases from the first 3 brightest lines \OIII, \OII to \Hb, convincing us of this weak correlation.
Linear fitting is employed in an attempt to describe this phenomenon, although it is based on limited data.
This non-zero inconsistency first appears when we use 1D \OIII flux to normalize our individual spectra for stacking.
We rechecked our MZR using 2D \OIII flux to normalize for stacking, and found the bias of metallicity is lower than $1\sigma$.
It indicates that the bias of the two flux measurements may be obscured by the stacking procedure, although we need larger a sample and more tests to verify this assertion. %in previous work(e.g., \tocite{???}).
A more significant effect may be seen in the physical quantities directly determined by the line flux value, such as SFR.

Since the flux ratio of 2D to 1D exhibits a correlation with the half-light radius $r_{50}$, we interpret this discrepancy as a morphological broadening effect.
The morphological broadening of the spectrum is not due to physical factors such as velocity dispersion or radiative damping, but is simply an observational effect of the extended source \citep{vanDokkum:2011cq, Wang:2020bp}.
For an ideal point source with no physical broadening effect, the emission line will be measured as a $\delta$ function.
But if we could spatially resolve the galaxy, which is common in slitless spectroscopy, the emission line would be broadened as a result of the superposition of $\delta$ functions from individual pixels.
Therefore, more parts of the line edge will be drowned in the noise, resulting in lower total line flux modeled by the Gaussian function.
And of course, larger sources produce more broadening, yielding lower flux measurements.
We, therefore, deem the top 3 panels of Figure~\ref{fig:compareflux} to be the first attempt to quantitatively analyze the impact of the morphological broadening effect.
For large sources ($r_{50}>10$), the intrinsic flux could be several times larger than the broadened flux.

Although the 2D measurements are larger than the 1D results, in general, it seems that this bias is the same for all emission lines of the same source.
As one can notice in the top 3 panels of Fig.~\ref{fig:compareflux}, for a given source with the same abscissa $r_{50}$, the corresponding ordinate values 2D/1D of all 3 lines are quite close to each other, although our naked eye can only recognize those outliers.
And we have tested that these patterns are also independent of their SNRs.
Moreover, we show the line flux ratio in the bottom 3 panels of Fig.\ref{fig:compareflux}, and they nearly follow the one-on-one line, with few outliers.
That means even if this effect is not taken into account like in the 1D method, the flux ratios do not deviate from the 2D method significantly.
Therefore, it indicates that the bias of the morphological broadening effect is systematic.
We color-code them with the metallicity or the dust extinction $A_v$
derived in Sect.\ref{subsec:metal} using 2D \grzl flux ratio.
The color patterns demonstrate the physical meaning of these line ratios, i.e., the gas-phase metallicity diagnostics $O_{32}:=$\OIII/\OII, $O_{3}:=$\OIII/\Hb, and the dust extinction indicator \Ha/\Hb.
The dotted line in the lower right marks the 'intrinsic' line ratio in the absence of dust attenuation $\Ha/\Hb=2.86$. 
The few sources below it may be due to low SNR and measurement errors \citep[see e.g.,][]{Nelson:2016ApJ...817L...9N}.

As a consequence, our key result of the metallicity measurement derived from the ratio of two lines in Sect.\ref{subsec:metal}, will not be greatly influenced by the 2D/1D flux measurement method.
However the direct line flux (e.g., \Hb) and the derived quantity (e.g., SFR) of a single emission line could be biased, and for a large source, the intrinsic flux could be several times larger than the measured one.
The coarse linear fitting here might describe the distinction between 2D/1D forward modeling flux of emission line to some extent.
We interpret this discrepancy as a morphological broadening effect.
We recommend carefully checking the way flux is measured to match the scientific requirement, and carefully forward modeling the spectrum through the convolution of the morphological broadening effect.
The systematic offset, for the first time, may present an important guideline for future work deriving line fluxes with wide-field slitless spectroscopy, especially for large sky surveys to be conducted by \eg, Euclid, Roman, and CSST, where it is time-consuming for 2D emission line modeling.

\section{Conclusions}\label{sec:conclude}

We have presented a comprehensive measurement of the MZR at a dwarf mass range using grism slitless spectroscopy.
The grism data are acquired by the GLASS-JWST ERS program, targeting the A2744 lensed field.
From the joint analysis of the JWST/NIRISS and
JWST/NIRCam data, we select a secure sample of 50 field galaxies with $M_*/M_\odot \in [10^{6.9},10^{10.0}] $ and \oh$\in[7.8,8.7]$ at 2 redshift range $z\in [1.8,2.3]$ and $z\in[2.6,3.4]$, assuming the strong line calibration of \citep{Bian:2018km}.
Our galaxies are divided into several mass bins and their spectra are stacked to increase the SNR.
Then we apply our forward modeling Bayesian metallicity inference method to the stacked line fluxes.
We derive the MZR in the A2744 lensed field as $12 + \log(\mathrm{O/H}) = \beta \times\log(M_*/10^8M_\odot)+ Z_8$ with $\beta=0.223 \pm 0.017$ and $\beta=0.294 \pm 0.010$ in these two redshift ranges $z_{\rm median}=1.90$ and $z_{\rm median}=2.88$, respectively, as well as a slight evolution: $\Delta Z_8=-0.11\pm0.02$, as presented in Tab.~\ref{tab:mzr} and Fig.~\ref{fig:mzr}.
Our MZRs have slopes that are consistent with those reported by \citet{Sanders:2021ga} at the higher mass end and similar redshifts, suggesting that gas outflow mechanisms with the same metal removal efficiency extend to the low-mass regime ($\lesssim10^9M_*$) at cosmic noon.
This $M_*$ scaling of metallicity is well reproduced by the FIRE simulations \citep{Ma:2016gw}. 

In addition, we assess the impact of the morphological broadening on emission line measurement by comparing two methods of using 2D forward modeling and line profile fitting to 1D extracted spectra. 
We show that ignoring the morphological broadening effect when deriving line fluxes from grism spectra results in a systematic reduction of flux by $\sim30\%$ on average.
The coarse linear fitting in Fig.~\ref{fig:compareflux} could characterize the impact of the morphological broadening effect on modeling the emission line flux to some extent.
The direct value (e.g., \Hb) and derived quantity (e.g., SFR)  of a single emission line flux could be biased, if one does not account for the galaxy morphology. However, this systematic effect does not significantly influence the line ratio and its derived quantities, e.g., metallicity, dust extinction, age, etc..
For this reason, we recommend careful inspection of the line modeling, especially for the next generation of large sky surveys, e.g., Euclid, Roman, and CSST.

\begin{acknowledgements}

We would like to thank the anonymous referee for the constructive comments that help us improve the clarity of this paper.
This paper is dedicated to the memory of our beloved colleague Mario Nonino who passed away prematurely. We miss him and are indebted to him for his countless contributions to the GLASS-JWST project.
This work is based on observations made with the NASA/ESA/CSA James Webb Space Telescope. 
The data were obtained from the Mikulski Archive for Space Telescopes at the Space Telescope Science Institute, which is operated by the Association of Universities for Research in Astronomy, Inc., under NASA contract NAS 5-03127 for JWST. 
These observations are associated with program JWST-ERS-1324.
We acknowledge financial support from NASA through grant JWST-ERS-1324.
X. H. thanks Xiaolei Meng, Lei Sun, and Lilan Yang for the useful discussion.
We thank the entire GLASS team that helped shape the manuscript.
X. W. is supported by the Fundamental Research Funds for the Central Universities, and the CAS Project for Young Scientists in Basic Research, Grant No. YSBR-062. 
This research is supported in part by the Australian Research Council Centre of Excellence for All Sky Astrophysics in 3 Dimensions (ASTRO 3D), through project number CE170100013. We acknowledge support from the INAF Large Grant 2022 “Extragalactic Surveys with JWST”  (PI Pentericci). B.M. is supported by an Australian Government Research Training Program (RTP) Scholarship. K.G. is supported by the Australian Research Council through the Discovery Early Career Researcher Award (DECRA) Fellowship (project number DE220100766) funded by the Australian Government.  

\facilities{JWST(NIRISS, NIRCame), HST (ACS, WFC3)}.

\software{
\grzl \citep{Brammer:2021df},
\bagp \citep{Carnall:2018gb},
\textsc{LMfit} \citep{Newville:2021cv},
\emc \citep{ForemanMackey:2013io}.
}

\end{acknowledgements}   

%= = = = = = = = = = = = = = = = = = = = = = = = = = = = = = = = = = = = = = = =
%= = = = = = = = = = = = = = = = = = = = = = = = = = = = = = = = = = = = = = = =
\appendix
\setcounter{table}{0}  
\renewcommand{\thetable}{A\arabic{table}}

\section{measured quantities of our sample}
In Tab.~\ref{tab:indi}, we show the observed and measured physical properties of all the 50 galaxies in our sample, including galaxy ID (ID Grism), coordinates (R.A. and Decl.) and grism redshift ($z_\mathrm{grizm}$) analyzed by \grzl, the matched ID in photometry of \cite{Paris:2023arXiv230102179P} (ID Photo.), the stellar mass $M_*$ estimated by SED fitting, the gravitational lensing magnification $\mu$ calculated using the model of \cite{Bergamini:2023arXiv230310210B}, and the dust attenuation ($A_\nu$), the de-redden Balmer emission line flux $f_{\Hb}$ (with its derived SFR), and the gas-phase metallicity \oh jointly estimated using our Bayesian method.
Note that $M_*, SFR$ have already been corrected by lensing magnification $\mu$, but $f_{\Hb}$ has not.
In Tab.~\ref{tab:flux}, we exhibit the emission line flux measurements by 2D/1D method, which are discussed in detail in Sect.~\ref{subsec:lineflux}.
Note that all $f_\mathrm{line}$ are not corrected by $\mu$.

\startlongtable

%\tabletypesize{\scriptsize}
%\tabcolsep=2pt
\begin{deluxetable*}{ccccccccccc}
%\tablenum{A1}   
\tablecaption{Measured Properties of Individual Galaxies\label{tab:indi}}
\tablewidth{0pt}
\tablehead{
\colhead{ID Grism} & \colhead{R.A.} & \colhead{Decl.} & \colhead{$z_{\rm grism}$} & \colhead{ID Photo.} & \colhead{log$(M_\star/M_\odot)$} & \colhead{$\mu$} & \colhead{$A_V$} & 
\colhead{deredden $f_{H\beta}$} & \colhead{SFR} &      
\colhead{12+ log(O/H)} \\
\colhead{} & \colhead{deg.} & \colhead{deg.} & \colhead{} & \colhead{} & \colhead{} & \colhead{} & \colhead{} & \colhead{$10^{-17}$\Funit} & \colhead{$\rm M_\odot\cdot yr^{-1}$} & \colhead{} 
}         
\decimalcolnumbers
\startdata
\multicolumn{11}{c}{$1.8<z_\mathrm{grism}<2.3$}\\
00765 & 3.5863967 & -30.4093408 & 2.014 & 04016 & $7.458_{-0.008}^{+0.008}$ & 4.97 & $0.45_{-0.10}^{+0.10}$ & $2.69_{-0.22}^{+0.24}$ & $2.12_{-0.17}^{+0.19}$ & $8.00_{-0.04}^{+0.06}$ \\ 
00902 & 3.6170966 & -30.4083725 & 1.876 & 04154 & $8.089_{-0.053}^{+0.041}$ & 1.68 & $0.14_{-0.11}^{+0.24}$ & $0.18_{-0.02}^{+0.05}$ & $0.36_{-0.05}^{+0.09}$ & $8.08_{-0.12}^{+0.10}$ \\ 
01331 & 3.5766842 & -30.4060897 & 1.808 & 04499 & $8.101_{-0.017}^{+0.016}$ & 2.37 & $0.70_{-0.28}^{+0.30}$ & $0.53_{-0.13}^{+0.18}$ & $0.68_{-0.16}^{+0.23}$ & $8.24_{-0.07}^{+0.06}$ \\ 
01365 & 3.6141973 & -30.4060640 & 2.275 & 04375 & $9.597_{-0.020}^{+0.029}$ & 1.77 & $0.10_{-0.07}^{+0.17}$ & $0.78_{-0.09}^{+0.14}$ & $2.32_{-0.27}^{+0.41}$ & $8.61_{-0.06}^{+0.05}$ \\ 
02128 & 3.5985318 & -30.4017605 & 2.009 & 05117 & $7.293_{-0.008}^{+0.009}$ & 3.55 & $0.21_{-0.10}^{+0.11}$ & $1.90_{-0.15}^{+0.18}$ & $2.08_{-0.17}^{+0.20}$ & $7.90_{-0.04}^{+0.04}$ \\ 
02332 & 3.6023553 & -30.4007355 & 1.804 & 05120 & $8.727_{-0.032}^{+0.020}$ & 2.55 & $0.05_{-0.04}^{+0.09}$ & $1.08_{-0.08}^{+0.11}$ & $1.26_{-0.09}^{+0.13}$ & $8.31_{-0.06}^{+0.05}$ \\ 
02696 & 3.6164407 & -30.3977732 & 1.996 & 05681 & $8.386_{-0.020}^{+0.022}$ & 1.65 & $0.11_{-0.08}^{+0.14}$ & $0.69_{-0.05}^{+0.09}$ & $1.59_{-0.11}^{+0.22}$ & $8.13_{-0.04}^{+0.04}$ \\ 
02793 & 3.6041872 & -30.3971816 & 2.068 & 05706 & $9.272_{-0.007}^{+0.006}$ & 2.10 & $0.65_{-0.15}^{+0.15}$ & $2.29_{-0.27}^{+0.30}$ & $4.53_{-0.54}^{+0.60}$ & $8.50_{-0.05}^{+0.04}$ \\ 
03393 & 3.6060398 & -30.3935272 & 2.177 & 06411 & $8.450_{-0.006}^{+0.007}$ & 1.93 & $0.34_{-0.21}^{+0.26}$ & $0.58_{-0.10}^{+0.16}$ & $1.41_{-0.25}^{+0.39}$ & $7.97_{-0.05}^{+0.05}$ \\ 
03557 & 3.6118150 & -30.3924863 & 2.278 & 06511 & $8.442_{-0.008}^{+0.006}$ & 1.75 & $0.01_{-0.01}^{+0.02}$ & $3.26_{-0.07}^{+0.08}$ & $9.84_{-0.20}^{+0.23}$ & $8.00_{-0.03}^{+0.04}$ \\ 
03666 & 3.6042544 & -30.3916573 & 1.880 & 06722 & $8.574_{-0.018}^{+0.018}$ & 1.91 & $0.80_{-0.24}^{+0.27}$ & $0.73_{-0.14}^{+0.19}$ & $1.26_{-0.24}^{+0.33}$ & $8.27_{-0.07}^{+0.06}$ \\ 
03784 & 3.6031401 & -30.3910461 & 2.177 & 06860 & $7.622_{-0.014}^{+0.013}$ & 1.99 & $0.04_{-0.03}^{+0.07}$ & $0.74_{-0.03}^{+0.05}$ & $1.75_{-0.06}^{+0.11}$ & $7.94_{-0.03}^{+0.04}$ \\ 
03785 & 3.6132702 & -30.3910937 & 1.879 & 06796 & $8.832_{-0.008}^{+0.008}$ & 1.67 & $1.02_{-0.11}^{+0.11}$ & $2.77_{-0.25}^{+0.27}$ & $5.47_{-0.49}^{+0.54}$ & $8.35_{-0.05}^{+0.04}$ \\ 
03854 & 3.5867377 & -30.3907657 & 2.206 & 06519 & $9.827_{-0.004}^{+0.003}$ & 6.37 & $1.47_{-0.09}^{+0.09}$ & $28.19_{-2.07}^{+2.24}$ & $21.60_{-1.59}^{+1.72}$ & $8.43_{-0.03}^{+0.03}$ \\ 
04001 & 3.6100121 & -30.3894795 & 2.173 & 07049 & $8.746_{-0.002}^{+0.002}$ & 1.73 & $0.14_{-0.10}^{+0.14}$ & $0.97_{-0.08}^{+0.12}$ & $2.64_{-0.22}^{+0.33}$ & $8.20_{-0.05}^{+0.06}$ \\ 
04457 & 3.5869454 & -30.3870037 & 1.858 & 07544 & $7.936_{-0.018}^{+0.020}$ & 3.60 & $0.14_{-0.10}^{+0.15}$ & $1.32_{-0.12}^{+0.19}$ & $1.17_{-0.11}^{+0.17}$ & $8.28_{-0.05}^{+0.04}$ \\ 
04482 & 3.5819407 & -30.3866370 & 1.884 & 07610 & $8.282_{-0.039}^{+0.037}$ & 3.69 & $0.49_{-0.30}^{+0.40}$ & $0.69_{-0.17}^{+0.30}$ & $0.62_{-0.15}^{+0.27}$ & $8.25_{-0.10}^{+0.08}$ \\ 
04539 & 3.5988518 & -30.3863743 & 1.857 & 07644 & $8.819_{-0.012}^{+0.012}$ & 2.09 & $0.15_{-0.10}^{+0.19}$ & $0.81_{-0.08}^{+0.14}$ & $1.24_{-0.12}^{+0.22}$ & $8.36_{-0.09}^{+0.07}$ \\ 
04579 & 3.5993864 & -30.3861434 & 2.060 & 07704 & $8.045_{-0.060}^{+0.047}$ & 2.07 & $0.32_{-0.22}^{+0.35}$ & $0.28_{-0.05}^{+0.11}$ & $0.56_{-0.11}^{+0.23}$ & $8.17_{-0.10}^{+0.09}$ \\ 
04611 & 3.5790397 & -30.3859412 & 2.187 & 07751 & $7.423_{-0.019}^{+0.021}$ & 3.65 & $0.04_{-0.03}^{+0.07}$ & $1.13_{-0.05}^{+0.07}$ & $1.48_{-0.06}^{+0.09}$ & $7.92_{-0.04}^{+0.04}$ \\ 
04842 & 3.5992144 & -30.3841762 & 2.028 & 08183 & $7.675_{-0.063}^{+0.048}$ & 2.00 & $0.31_{-0.23}^{+0.47}$ & $0.18_{-0.04}^{+0.10}$ & $0.35_{-0.08}^{+0.20}$ & $7.96_{-0.09}^{+0.09}$ \\ 
04946 & 3.5701934 & -30.3837325 & 1.860 & 08099 & $9.245_{-0.004}^{+0.004}$ & 2.96 & $0.81_{-0.09}^{+0.10}$ & $6.44_{-0.50}^{+0.55}$ & $6.98_{-0.54}^{+0.59}$ & $8.45_{-0.04}^{+0.04}$ \\ 
05123 & 3.5920216 & -30.3825005 & 1.860 & 08565 & $6.894_{-0.022}^{+0.021}$ & 2.52 & $1.16_{-0.27}^{+0.27}$ & $3.38_{-0.68}^{+0.85}$ & $4.31_{-0.86}^{+1.08}$ & $7.99_{-0.08}^{+0.08}$ \\ 
05715 & 3.6103731 & -30.3801845 & 1.877 & 08541 & $9.688_{-0.013}^{+0.017}$ & 1.75 & $0.07_{-0.05}^{+0.10}$ & $1.66_{-0.08}^{+0.14}$ & $3.12_{-0.15}^{+0.26}$ & $8.55_{-0.05}^{+0.04}$ \\ 
05747 & 3.5985949 & -30.3785188 & 1.915 & 09272 & $9.087_{-0.003}^{+0.003}$ & 1.97 & $0.07_{-0.06}^{+0.12}$ & $1.29_{-0.07}^{+0.14}$ & $2.26_{-0.13}^{+0.24}$ & $8.21_{-0.09}^{+0.09}$ \\ 
05770 & 3.5997721 & -30.3778656 & 1.880 & 09586 & $8.124_{-0.007}^{+0.008}$ & 1.94 & $0.52_{-0.25}^{+0.29}$ & $0.50_{-0.10}^{+0.15}$ & $0.86_{-0.18}^{+0.26}$ & $7.91_{-0.06}^{+0.06}$ \\ 
05866 & 3.5911011 & -30.3816997 & 1.883 & 08556 & $8.764_{-0.002}^{+0.002}$ & 2.53 & $0.36_{-0.07}^{+0.07}$ & $7.80_{-0.44}^{+0.48}$ & $10.20_{-0.58}^{+0.63}$ & $8.12_{-0.04}^{+0.04}$ \\ 
05952 & 3.5950311 & -30.3761179 & 1.832 & 09990 & $7.142_{-0.023}^{+0.022}$ & 2.06 & $1.23_{-0.21}^{+0.21}$ & $9.04_{-1.48}^{+1.73}$ & $13.57_{-2.22}^{+2.59}$ & $8.09_{-0.06}^{+0.06}$ \\ 
\hline
\multicolumn{11}{c}{$2.6<z_\mathrm{grism}<3.4$}\\
00073 & 3.5893372 & -30.4159113 & 2.647 & 02987 & $9.594_{-0.002}^{+0.002}$ & 3.02 & $0.04_{-0.03}^{+0.07}$ & $0.94_{-0.07}^{+0.09}$ & $2.36_{-0.18}^{+0.24}$ & $8.50_{-0.04}^{+0.04}$ \\ 
00671 & 3.5845970 & -30.4097995 & 2.657 & 03939 & $8.673_{-0.003}^{+0.003}$ & 3.94 & $0.30_{-0.22}^{+0.35}$ & $1.80_{-0.41}^{+0.91}$ & $3.50_{-0.79}^{+1.78}$ & $8.23_{-0.06}^{+0.07}$ \\ 
01192 & 3.6134541 & -30.4068477 & 2.848 & 04407 & $7.536_{-0.037}^{+0.051}$ & 1.87 & $3.46_{-0.82}^{+0.38}$ & $14.35_{-9.01}^{+8.19}$ & $69.48_{-43.64}^{+39.67}$ & $8.32_{-0.13}^{+0.09}$ \\ 
01514 & 3.6074237 & -30.4064785 & 3.196 & 04281 & $9.358_{-0.002}^{+0.002}$ & 2.47 & $0.31_{-0.21}^{+0.32}$ & $1.49_{-0.34}^{+0.70}$ & $7.21_{-1.65}^{+3.37}$ & $8.33_{-0.06}^{+0.06}$ \\ 
01588 & 3.6129938 & -30.4050844 & 3.043 & 04550 & $8.844_{-0.011}^{+0.010}$ & 1.87 & $1.95_{-0.43}^{+0.38}$ & $30.17_{-11.84}^{+17.12}$ & $171.88_{-67.45}^{+97.54}$ & $8.45_{-0.06}^{+0.05}$ \\ 
01589 & 3.6128172 & -30.4049834 & 3.042 & 04444 & $9.489_{-0.013}^{+0.012}$ & 1.88 & $1.60_{-0.45}^{+0.41}$ & $18.68_{-7.79}^{+11.97}$ & $105.86_{-44.17}^{+67.84}$ & $8.58_{-0.06}^{+0.05}$ \\ 
01659 & 3.6198203 & -30.4043177 & 2.922 & 04709 & $8.984_{-0.003}^{+0.003}$ & 1.72 & $0.34_{-0.24}^{+0.42}$ & $2.34_{-0.52}^{+1.33}$ & $13.11_{-2.91}^{+7.44}$ & $7.92_{-0.04}^{+0.05}$ \\ 
02025 & 3.5982393 & -30.4023120 & 2.651 & 04978 & $8.593_{-0.002}^{+0.002}$ & 4.30 & $1.50_{-0.41}^{+0.36}$ & $7.80_{-2.95}^{+3.96}$ & $13.84_{-5.22}^{+7.03}$ & $8.43_{-0.07}^{+0.05}$ \\ 
02389 & 3.6094671 & -30.4003762 & 2.665 & 05237 & $9.550_{-0.013}^{+0.003}$ & 1.94 & $0.18_{-0.13}^{+0.24}$ & $2.01_{-0.31}^{+0.67}$ & $7.99_{-1.22}^{+2.68}$ & $8.33_{-0.05}^{+0.05}$ \\ 
02621 & 3.6136448 & -30.3986436 & 2.843 & 05484 & $9.252_{-0.007}^{+0.008}$ & 1.80 & $0.16_{-0.12}^{+0.22}$ & $1.67_{-0.30}^{+0.53}$ & $8.35_{-1.50}^{+2.67}$ & $8.58_{-0.06}^{+0.06}$ \\ 
02654 & 3.6118526 & -30.3981734 & 3.041 & 05594 & $8.928_{-0.012}^{+0.013}$ & 1.89 & $0.33_{-0.25}^{+0.51}$ & $0.49_{-0.14}^{+0.44}$ & $2.74_{-0.80}^{+2.49}$ & $8.43_{-0.08}^{+0.08}$ \\ 
02703 & 3.6093784 & -30.3983894 & 2.691 & 05425 & $9.975_{-0.002}^{+0.002}$ & 1.94 & $0.79_{-0.37}^{+0.37}$ & $8.82_{-3.01}^{+4.61}$ & $35.95_{-12.26}^{+18.77}$ & $8.57_{-0.04}^{+0.04}$ \\ 
02855 & 3.5749452 & -30.3967746 & 3.125 & 05793 & $8.053_{-0.034}^{+0.035}$ & 3.71 & $2.84_{-0.95}^{+0.73}$ & $51.19_{-33.57}^{+65.75}$ & $156.43_{-102.59}^{+200.91}$ & $8.10_{-0.11}^{+0.10}$ \\ 
02913 & 3.6078376 & -30.3962862 & 2.666 & 05857 & $8.679_{-0.009}^{+0.008}$ & 2.12 & $0.36_{-0.26}^{+0.50}$ & $0.52_{-0.15}^{+0.49}$ & $1.90_{-0.55}^{+1.78}$ & $8.39_{-0.07}^{+0.08}$ \\ 
03018 & 3.6070933 & -30.3956151 & 2.980 & 06039 & $7.838_{-0.006}^{+0.006}$ & 2.10 & $0.79_{-0.33}^{+0.38}$ & $2.75_{-0.85}^{+1.41}$ & $13.27_{-4.10}^{+6.78}$ & $7.82_{-0.02}^{+0.03}$ \\ 
03531 & 3.6112440 & -30.3924593 & 2.981 & 06626 & $7.154_{-0.035}^{+0.037}$ & 1.81 & $2.65_{-1.13}^{+0.94}$ & $4.85_{-3.53}^{+9.43}$ & $27.17_{-19.77}^{+52.86}$ & $7.85_{-0.04}^{+0.12}$ \\ 
04898 & 3.6022598 & -30.3843036 & 2.663 & 07846 & $9.904_{-0.003}^{+0.004}$ & 1.96 & $0.59_{-0.35}^{+0.40}$ & $5.77_{-1.99}^{+3.50}$ & $22.71_{-7.84}^{+13.78}$ & $8.38_{-0.08}^{+0.06}$ \\ 
05184 & 3.5859437 & -30.3821176 & 3.053 & 08570 & $7.880_{-0.008}^{+0.007}$ & 3.28 & $1.19_{-0.44}^{+0.51}$ & $4.26_{-1.65}^{+3.33}$ & $13.95_{-5.39}^{+10.91}$ & $8.10_{-0.06}^{+0.07}$ \\ 
05343 & 3.5778395 & -30.3811884 & 3.390 & 08654 & $9.129_{-0.008}^{+0.008}$ & 3.45 & $0.14_{-0.11}^{+0.22}$ & $2.45_{-0.31}^{+0.71}$ & $9.80_{-1.23}^{+2.84}$ & $8.25_{-0.04}^{+0.05}$ \\ 
05475 & 3.6060732 & -30.3801651 & 2.691 & 08838 & $9.951_{-0.003}^{+0.003}$ & 1.82 & $0.42_{-0.26}^{+0.32}$ & $8.46_{-2.21}^{+3.81}$ & $36.75_{-9.62}^{+16.57}$ & $8.56_{-0.05}^{+0.04}$ \\ 
05526 & 3.5914083 & -30.3797763 & 2.718 & 08958 & $8.432_{-0.003}^{+0.004}$ & 2.53 & $0.07_{-0.05}^{+0.11}$ & $4.10_{-0.24}^{+0.51}$ & $13.12_{-0.78}^{+1.63}$ & $8.08_{-0.04}^{+0.04}$ \\ 
06057 & 3.6033100 & -30.3742575 & 3.043 & 10305 & $9.083_{-0.006}^{+0.006}$ & 1.84 & $0.94_{-0.63}^{+0.73}$ & $2.58_{-1.42}^{+4.15}$ & $14.86_{-8.17}^{+23.91}$ & $8.35_{-0.10}^{+0.11}$ \\  
\enddata

\tablecomments{Column 1 is the source ID reduced from JWST/NIRISS Grism data by our source detection \grzl procedure; Columns 2 and 3 are the equatorial coordinates right ascension (R.A.) and declination (Decl.) in equinox with an epoch of J2000; Column 4 is the secure redshift determined by \grzl in Sec.\ref{subsec:gri}.; Column 5 is the matched ID of the GLASS photometric catalog \citet{Paris:2023arXiv230102179P};
Column 6 is the stellar mass fitted from the catalog; Column 7 is the magnification of the gravitational lensing effect by the Abell 2744 cluster. 
Column 8,9 is the dust attenuation $A_V$ \& de-redden $H_\beta$ flux estimated in Sec.\ref{subsec:metal}; Column 10 is the star formation rate determined by de-redden $f_{H_\beta}$;
Column 11 is gas phase metallicity represented by oxygen abundance.}
\end{deluxetable*}

%= = = = = = = = = = = = = = = = = = = = = = = = = = = = = = = = = = = = = = = =
\begin{longrotatetable}
% = = = = = = = = = = = = = = = = = = = = = = = = = = = = = = = = = = = = = = = = = =
% Include this table with \input{filename.tex}
% To rotate in emulateapj do: \begin{turnpage}\input{filename.tex}\end{turnpage}
% To display it on multiple pages do: \LongTables\input{filename.tex}
% \multicolumn{6}{c}{\hrulefill}, \hrulefill means a horizontal line, as a row.
% - - - - - - - - - - - - - - - - - - - - - - - - - - - - - - - - - - - - - - - - - -
{
\tabletypesize{\scriptsize} % table font size, \scriptsize=7pt
\tabcolsep=3pt   % separation of columns 
\begin{deluxetable*}{cccccccccccccccccc}   \tablecolumns{18}
\tablewidth{1pt}
\tablecaption{Flux derived from 2D/1D forward modeling of the individual galaxies.}
\label{tab:flux}
% - - - - - - - - - - - - - - - - - - - - - - - - - - - - - - - - - - - - - - - - - -
\tablehead{
    \colhead{ID} & 
    \colhead{R.A.} & 
    \colhead{Dec.} & 
    \colhead{$z_\mathrm{grism}$} & 
    \multicolumn{7}{c}{2D forward modeling of emission line fluxes $f_\mathrm{line}$ [$10^{-17}$\Funit]} &
    \multicolumn{7}{c}{1D extracted line profile fitting of emission line fluxes $f_\mathrm{line}$ [$10^{-17}$\Funit]} \\
    % & [deg.] & [deg.] & &  \multicolumn{6}{c}{\hrulefill} & \multicolumn{6}{c}{\hrulefill} \\
    \cline{5-11} \cline{12-18}
    & [deg.] & [deg.] & &  
    %\colhead{$f_{\rm [OII]}$} & \colhead{$f_{\rm [NeIII]}$} & \colhead{$f_{\rm H\gamma}$} & \colhead{$f_{\rm H\beta}$} & \colhead{$f_{\rm [OIII]}$} & \colhead{$f_{\rm H\alpha}$} & \colhead{$f_{\rm [SII]}$} & 
    \colhead{${\rm [OII]}$} & 
    \colhead{${\rm [NeIII]}$} & 
    \colhead{${\rm H\gamma}$} & 
    \colhead{${\rm H\beta}$} & 
    \colhead{${\rm [OIII]}$} & 
    \colhead{${\rm H\alpha}$} &
    \colhead{${\rm [SII]}$} &
    \colhead{${\rm [OII]}$} & 
    \colhead{${\rm [NeIII]}$} & 
    \colhead{${\rm H\gamma}$} & 
    \colhead{${\rm H\beta}$} & 
    \colhead{${\rm [OIII]}$} & 
    \colhead{${\rm H\alpha}$} &
    \colhead{${\rm [SII]}$} 
}
% - - - - - - - - - - - - - - - - - - - - - - - - - - - - - - - - - - - - - - - - - -
\startdata
    \noalign{\smallskip}\hline\noalign{\smallskip}
    \multicolumn{18}{c}{$1.8<z_\mathrm{grism}<2.6$}\\
    \noalign{\smallskip}
00765 & 3.5863967 & -30.4093408 & 2.014 & $1.44\pm0.07$ & $0.69\pm0.09$ & ... & $1.69\pm0.05$ & $14.92\pm0.10$ & $5.49\pm0.09$ & $0.77\pm0.08$ & $1.52\pm0.15$ & $1.04\pm0.11$ & ... & $1.84\pm0.10$ & $14.51\pm0.19$ & $4.77\pm0.16$ & $0.79\pm0.16$ \\
00902 & 3.6170966 & -30.4083725 & 1.876 & $0.37\pm0.10$ & $0.11\pm0.12$ & $0.17\pm0.07$ & $0.20\pm0.05$ & $1.82\pm0.09$ & $0.35\pm0.07$ & $0.11\pm0.07$ & $0.11\pm0.08$ & $0.01\pm0.06$ & ... & $0.08\pm0.04$ & $0.66\pm0.07$ & $0.14\pm0.06$ & $0.02\pm0.05$ \\
01331 & 3.5766842 & -30.4060897 & 1.808 & $0.42\pm0.05$ & $0.07\pm0.06$ & $0.10\pm0.03$ & $0.25\pm0.05$ & $2.31\pm0.08$ & $0.89\pm0.07$ & $0.21\pm0.08$ & $0.33\pm0.06$ & $0.07\pm0.04$ & ... & $0.18\pm0.05$ & $1.42\pm0.07$ & $0.60\pm0.07$ & $0.19\pm0.08$ \\
01365 & 3.6141973 & -30.4060640 & 2.275 & $2.00\pm0.20$ & $0.86\pm0.42$ & $0.02\pm0.35$ & $0.61\pm0.17$ & $1.72\pm0.23$ & $1.96\pm0.22$ & $0.77\pm0.23$ & $0.67\pm0.13$ & ... & ... & $0.24\pm0.13$ & $0.73\pm0.21$ & $0.74\pm0.16$ & $0.42\pm0.18$ \\
02128 & 3.5985318 & -30.4017605 & 2.009 & $1.04\pm0.06$ & $0.71\pm0.08$ & ... & $1.52\pm0.05$ & $12.32\pm0.09$ & $4.65\pm0.08$ & $0.19\pm0.07$ & $1.14\pm0.15$ & $0.84\pm0.16$ & ... & $1.58\pm0.12$ & $10.95\pm0.21$ & $4.15\pm0.22$ & $0.17\pm0.17$ \\
02332 & 3.6023553 & -30.4007355 & 1.804 & $3.37\pm0.40$ & $1.28\pm1.12$ & ... & $1.31\pm0.16$ & $8.22\pm0.23$ & $2.35\pm0.24$ & ... & $0.95\pm0.21$ & ... & ... & $0.40\pm0.11$ & $2.60\pm0.16$ & $0.72\pm0.16$ & ... \\
02696 & 3.6164407 & -30.3977732 & 1.996 & $1.10\pm0.07$ & $0.21\pm0.08$ & ... & $0.59\pm0.05$ & $6.98\pm0.09$ & $1.77\pm0.07$ & $0.08\pm0.08$ & $0.71\pm0.07$ & $0.14\pm0.06$ & ... & $0.25\pm0.05$ & $3.93\pm0.09$ & $1.04\pm0.08$ & $0.04\pm0.05$ \\
02793 & 3.6041872 & -30.3971816 & 2.068 & $2.41\pm0.10$ & $0.15\pm0.22$ & $0.03\pm0.27$ & $1.08\pm0.07$ & $5.30\pm0.11$ & $4.07\pm0.12$ & $0.71\pm0.12$ & $1.77\pm0.14$ & ... & $0.38\pm0.13$ & $0.74\pm0.11$ & $4.09\pm0.17$ & $3.07\pm0.16$ & $0.37\pm0.16$ \\
03393 & 3.6060398 & -30.3935272 & 2.177 & $0.36\pm0.03$ & $0.23\pm0.05$ & $0.10\pm0.06$ & $0.39\pm0.05$ & $5.21\pm0.09$ & $1.26\pm0.08$ & $0.15\pm0.08$ & $0.24\pm0.03$ & ... & ... & $0.20\pm0.04$ & $3.02\pm0.07$ & $0.69\pm0.07$ & $0.11\pm0.06$ \\
03557 & 3.6118150 & -30.3924863 & 2.278 & $3.72\pm0.13$ & $1.31\pm0.20$ & $1.64\pm0.20$ & $4.33\pm0.15$ & $40.62\pm0.28$ & $8.57\pm0.18$ & ... & $1.11\pm0.36$ & ... & $0.54\pm0.29$ & $2.47\pm0.31$ & $14.33\pm0.66$ & $2.27\pm0.38$ & ... \\
03666 & 3.6042544 & -30.3916573 & 1.880 & $0.50\pm0.06$ & $0.13\pm0.07$ & $0.12\pm0.04$ & $0.33\pm0.05$ & $2.51\pm0.07$ & $1.14\pm0.06$ & $0.14\pm0.06$ & $0.40\pm0.08$ & $0.12\pm0.06$ & $0.14\pm0.04$ & $0.26\pm0.06$ & $1.86\pm0.08$ & $0.99\pm0.06$ & $0.17\pm0.07$ \\
03784 & 3.6031401 & -30.3910461 & 2.177 & $0.62\pm0.04$ & $0.33\pm0.05$ & $0.34\pm0.05$ & $0.85\pm0.04$ & $8.27\pm0.08$ & $1.95\pm0.06$ & $0.04\pm0.06$ & $0.41\pm0.06$ & $0.29\pm0.06$ & $0.31\pm0.05$ & $0.52\pm0.06$ & $5.53\pm0.10$ & $1.31\pm0.08$ & $0.07\pm0.05$ \\
03785 & 3.6132702 & -30.3910937 & 1.879 & $1.80\pm0.08$ & $0.24\pm0.13$ & $0.27\pm0.07$ & $0.90\pm0.05$ & $7.12\pm0.08$ & $3.71\pm0.07$ & $0.57\pm0.07$ & $1.39\pm0.11$ & $0.42\pm0.09$ & $0.21\pm0.07$ & $0.55\pm0.07$ & $5.06\pm0.11$ & $2.77\pm0.09$ & $0.31\pm0.10$ \\
03854 & 3.5867377 & -30.3907657 & 2.206 & $16.62\pm0.53$ & ... & $3.90\pm0.76$ & $2.22\pm0.32$ & $28.81\pm0.43$ & $27.59\pm0.47$ & $2.21\pm0.47$ & $5.87\pm0.56$ & ... & ... & $1.26\pm0.51$ & $9.92\pm0.88$ & $12.02\pm0.75$ & $1.90\pm0.72$ \\
04001 & 3.6100121 & -30.3894795 & 2.173 & $1.45\pm0.06$ & $0.19\pm0.08$ & $0.42\pm0.07$ & $0.90\pm0.06$ & $7.27\pm0.10$ & $2.45\pm0.09$ & $0.24\pm0.08$ & $0.81\pm0.06$ & $0.21\pm0.05$ & $0.23\pm0.05$ & $0.48\pm0.05$ & $3.95\pm0.09$ & $1.31\pm0.08$ & ... \\
04457 & 3.5869454 & -30.3870037 & 1.858 & $2.67\pm0.17$ & $0.78\pm0.35$ & $0.42\pm0.16$ & $1.04\pm0.08$ & $10.02\pm0.13$ & $3.37\pm0.16$ & $0.66\pm0.12$ & $1.29\pm0.23$ & $0.65\pm0.20$ & $0.12\pm0.11$ & $0.60\pm0.12$ & $5.68\pm0.19$ & $1.88\pm0.23$ & $0.30\pm0.17$ \\
04482 & 3.5819407 & -30.3866370 & 1.884 & $0.76\pm0.20$ & ... & $0.23\pm0.14$ & $0.43\pm0.08$ & $3.43\pm0.12$ & $1.35\pm0.13$ & ... & $0.36\pm0.12$ & $0.13\pm0.10$ & ... & $0.42\pm0.06$ & $2.09\pm0.11$ & $0.62\pm0.10$ & ... \\
04539 & 3.5988518 & -30.3863743 & 1.857 & $1.10\pm0.15$ & ... & $0.35\pm0.14$ & $0.91\pm0.07$ & $4.48\pm0.10$ & $1.94\pm0.11$ & $0.32\pm0.10$ & $0.52\pm0.16$ & ... & ... & $0.57\pm0.08$ & $3.09\pm0.13$ & $1.13\pm0.13$ & $0.20\pm0.12$ \\
04579 & 3.5993864 & -30.3861434 & 2.060 & $0.28\pm0.06$ & $0.20\pm0.08$ & $0.15\pm0.15$ & $0.36\pm0.05$ & $1.79\pm0.08$ & $0.59\pm0.07$ & $0.13\pm0.07$ & $0.27\pm0.08$ & $0.07\pm0.04$ & $0.14\pm0.06$ & $0.27\pm0.05$ & $1.28\pm0.09$ & $0.41\pm0.08$ & $0.16\pm0.09$ \\
04611 & 3.5790397 & -30.3859412 & 2.187 & $0.85\pm0.07$ & $0.66\pm0.12$ & $0.63\pm0.09$ & $1.31\pm0.07$ & $10.26\pm0.12$ & $2.93\pm0.12$ & ... & $0.49\pm0.10$ & $0.20\pm0.10$ & $0.32\pm0.09$ & $0.74\pm0.09$ & $5.89\pm0.16$ & $1.66\pm0.15$ & ... \\
04842 & 3.5992144 & -30.3841762 & 2.028 & $0.14\pm0.05$ & $0.07\pm0.06$ & ... & $0.09\pm0.06$ & $1.67\pm0.10$ & $0.31\pm0.08$ & $0.05\pm0.09$ & $0.04\pm0.04$ & ... & ... & $0.03\pm0.05$ & $0.61\pm0.06$ & $0.14\pm0.05$ & ... \\
04946 & 3.5701934 & -30.3837325 & 1.860 & $6.04\pm0.20$ & $1.46\pm0.32$ & $0.93\pm0.19$ & $2.56\pm0.10$ & $15.20\pm0.16$ & $10.09\pm0.18$ & $3.20\pm0.24$ & $4.40\pm0.41$ & $0.64\pm0.32$ & $0.50\pm0.28$ & $2.20\pm0.24$ & $11.87\pm0.39$ & $7.47\pm0.43$ & $1.95\pm0.57$ \\
05123 & 3.5920216 & -30.3825005 & 1.860 & $0.70\pm0.18$ & $0.80\pm0.23$ & $0.41\pm0.13$ & $0.98\pm0.12$ & $11.33\pm0.20$ & $4.06\pm0.17$ & ... & $0.33\pm0.09$ & $0.28\pm0.08$ & $0.13\pm0.05$ & $0.29\pm0.05$ & $2.84\pm0.10$ & $0.97\pm0.08$ & $0.05\pm0.05$ \\
05715 & 3.6103731 & -30.3801845 & 1.877 & $2.99\pm0.26$ & $0.24\pm0.52$ & ... & $1.82\pm0.11$ & $6.09\pm0.15$ & $4.36\pm0.13$ & $0.30\pm0.16$ & $1.41\pm0.30$ & ... & $0.17\pm0.18$ & $1.38\pm0.16$ & $4.56\pm0.24$ & $3.10\pm0.25$ & $0.53\pm0.24$ \\
05747 & 3.5985949 & -30.3785188 & 1.915 & $1.82\pm0.18$ & ... & $0.26\pm0.17$ & $1.69\pm0.10$ & $8.82\pm0.15$ & $3.29\pm0.13$ & $0.24\pm0.13$ & $1.23\pm0.17$ & $0.41\pm0.14$ & $0.20\pm0.10$ & $1.08\pm0.10$ & $5.16\pm0.16$ & $1.71\pm0.13$ & $0.22\pm0.13$ \\
05770 & 3.5997721 & -30.3778656 & 1.880 & $0.20\pm0.05$ & $0.05\pm0.04$ & $0.12\pm0.03$ & $0.32\pm0.04$ & $3.74\pm0.07$ & $0.94\pm0.06$ & ... & $0.07\pm0.02$ & ... & $0.03\pm0.02$ & $0.17\pm0.04$ & $1.86\pm0.06$ & $0.45\pm0.06$ & ... \\
05866 & 3.5911011 & -30.3816997 & 1.883 & $7.90\pm0.28$ & $1.44\pm0.52$ & $2.12\pm0.23$ & $5.38\pm0.12$ & $58.64\pm0.23$ & $17.07\pm0.17$ & $1.58\pm0.16$ & $4.25\pm1.03$ & $2.19\pm0.88$ & $0.51\pm0.60$ & $4.64\pm0.53$ & $30.49\pm0.98$ & $8.14\pm0.71$ & $1.12\pm0.66$ \\
05952 & 3.5950311 & -30.3761179 & 1.832 & $2.28\pm0.27$ & $0.44\pm0.36$ & $0.92\pm0.26$ & $2.63\pm0.21$ & $24.84\pm0.40$ & $10.22\pm0.36$ & ... & $0.47\pm0.13$ & $0.47\pm0.12$ & $0.18\pm0.10$ & $0.51\pm0.11$ & $4.60\pm0.18$ & $1.90\pm0.21$ & ... \\
    \noalign{\smallskip}\hline\noalign{\smallskip}
    \hline
    \multicolumn{18}{c}{$2.6<z_\mathrm{grism}<3.4$}\\
    \noalign{\smallskip}
00073 & 3.5893372 & -30.4159113 & 2.647 & $2.92\pm0.09$ & $0.68\pm0.16$ & $0.43\pm0.14$ & $0.52\pm0.11$ & $3.07\pm0.17$ & ... & ... & $1.85\pm0.18$ & $0.78\pm0.18$ & ... & $0.25\pm0.24$ & $1.25\pm0.34$ & ... & ... \\
00671 & 3.5845970 & -30.4097995 & 2.657 & $2.38\pm0.07$ & $0.67\pm0.12$ & $0.53\pm0.10$ & $1.32\pm0.11$ & $11.72\pm0.19$ & ... & ... & $1.38\pm0.06$ & $0.46\pm0.06$ & $0.39\pm0.07$ & $0.83\pm0.11$ & $6.44\pm0.17$ & ... & ... \\
01192 & 3.6134541 & -30.4068477 & 2.848 & $0.21\pm0.03$ & $0.05\pm0.03$ & ... & $0.39\pm0.04$ & $2.79\pm0.08$ & ... & ... & $0.31\pm0.06$ & $0.12\pm0.06$ & ... & $0.55\pm0.09$ & $2.93\pm0.15$ & ... & ... \\
01514 & 3.6074237 & -30.4064785 & 3.196 & $2.30\pm0.07$ & $0.17\pm0.13$ & $0.78\pm0.16$ & $0.95\pm0.11$ & $8.07\pm0.18$ & ... & ... & $1.79\pm0.09$ & $0.10\pm0.05$ & $0.60\pm0.12$ & $0.52\pm0.09$ & $5.94\pm0.19$ & ... & ... \\
01588 & 3.6129938 & -30.4050844 & 3.043 & $4.43\pm0.18$ & $0.33\pm0.37$ & $0.93\pm0.65$ & $3.78\pm0.27$ & $22.23\pm0.45$ & ... & ... & $1.56\pm0.14$ & $0.22\pm0.11$ & ... & $1.55\pm0.17$ & $7.23\pm0.31$ & ... & ... \\
01589 & 3.6128172 & -30.4049834 & 3.042 & $3.91\pm0.25$ & $0.39\pm0.42$ & ... & $3.47\pm0.40$ & $13.46\pm0.66$ & ... & ... & $0.90\pm0.11$ & $0.26\pm0.11$ & ... & $0.92\pm0.16$ & $3.18\pm0.26$ & ... & ... \\
01659 & 3.6198203 & -30.4043177 & 2.922 & $1.17\pm0.06$ & $0.79\pm0.10$ & $56.41\pm184.72$ & $1.67\pm0.09$ & $18.21\pm0.23$ & ... & ... & $0.70\pm0.08$ & $0.51\pm0.10$ & ... & $0.81\pm0.10$ & $10.26\pm0.26$ & ... & ... \\
02025 & 3.5982393 & -30.4023120 & 2.651 & $1.95\pm0.05$ & $0.46\pm0.09$ & $0.56\pm0.08$ & $1.59\pm0.08$ & $9.59\pm0.14$ & ... & ... & $2.18\pm0.11$ & $0.93\pm0.11$ & $0.40\pm0.13$ & $1.83\pm0.18$ & $8.63\pm0.28$ & ... & ... \\
02389 & 3.6094671 & -30.4003762 & 2.665 & $4.12\pm0.10$ & $0.46\pm0.22$ & $0.53\pm0.16$ & $1.46\pm0.16$ & $13.24\pm0.24$ & ... & ... & $2.85\pm0.18$ & $0.82\pm0.19$ & ... & $1.11\pm0.32$ & $8.52\pm0.47$ & ... & ... \\
02621 & 3.6136448 & -30.3986436 & 2.843 & $3.12\pm0.18$ & $0.78\pm0.49$ & $0.21\pm0.54$ & $1.24\pm0.25$ & $4.72\pm0.36$ & ... & ... & $1.08\pm0.12$ & $0.31\pm0.12$ & ... & $0.49\pm0.16$ & $1.66\pm0.25$ & ... & ... \\
02654 & 3.6118526 & -30.3981734 & 3.041 & $0.82\pm0.11$ & $0.21\pm0.12$ & ... & $0.16\pm0.12$ & $1.99\pm0.19$ & ... & ... & $0.40\pm0.06$ & $0.17\pm0.07$ & ... & $0.10\pm0.09$ & $0.83\pm0.14$ & ... & ... \\
02703 & 3.6093784 & -30.3983894 & 2.691 & $5.58\pm0.19$ & $0.80\pm0.28$ & $1.81\pm0.25$ & $3.77\pm0.22$ & $15.56\pm0.43$ & ... & ... & $1.80\pm0.16$ & $0.09\pm0.12$ & $0.35\pm0.13$ & $1.10\pm0.22$ & $4.42\pm0.28$ & ... & ... \\
02855 & 3.5749452 & -30.3967746 & 3.125 & $1.39\pm0.14$ & $0.44\pm0.19$ & $0.26\pm0.56$ & $2.65\pm0.34$ & $27.32\pm0.60$ & ... & ... & $0.56\pm0.11$ & ... & $0.48\pm0.31$ & $1.25\pm0.22$ & $9.57\pm0.37$ & ... & ... \\
02913 & 3.6078376 & -30.3962862 & 2.666 & $0.81\pm0.07$ & $0.23\pm0.12$ & $0.08\pm0.11$ & $0.13\pm0.12$ & $2.37\pm0.17$ & ... & ... & $0.52\pm0.08$ & $0.12\pm0.08$ & $0.22\pm0.08$ & $0.14\pm0.06$ & $1.43\pm0.18$ & ... & ... \\
03018 & 3.6070933 & -30.3956151 & 2.980 & $0.45\pm0.04$ & $0.44\pm0.05$ & $6.49\pm7.83$ & $1.24\pm0.06$ & $12.21\pm0.14$ & ... & ... & $0.49\pm0.08$ & $0.55\pm0.10$ & ... & $0.88\pm0.13$ & $7.95\pm0.22$ & ... & ... \\
03531 & 3.6112440 & -30.3924593 & 2.981 & $0.03\pm0.03$ & $0.09\pm0.03$ & ... & $0.52\pm0.06$ & $2.26\pm0.10$ & ... & ... & $0.04\pm0.02$ & $0.04\pm0.02$ & ... & $0.25\pm0.04$ & $0.92\pm0.07$ & ... & ... \\
04898 & 3.6022598 & -30.3843036 & 2.663 & $5.35\pm0.12$ & $1.11\pm0.24$ & $1.34\pm0.18$ & $3.15\pm0.23$ & $20.20\pm0.40$ & ... & ... & $2.27\pm0.10$ & $0.53\pm0.10$ & $0.34\pm0.12$ & $1.28\pm0.23$ & $7.77\pm0.39$ & ... & ... \\
05184 & 3.5859437 & -30.3821176 & 3.053 & $1.21\pm0.04$ & $0.40\pm0.06$ & $0.38\pm0.12$ & $1.19\pm0.07$ & $13.50\pm0.15$ & ... & ... & $0.89\pm0.07$ & $0.26\pm0.07$ & $0.14\pm0.10$ & $0.93\pm0.12$ & $10.98\pm0.21$ & ... & ... \\
05343 & 3.5778395 & -30.3811884 & 3.390 & $4.62\pm0.13$ & $6.91\pm71.99$ & $2.22\pm0.36$ & $1.58\pm0.20$ & $19.79\pm0.34$ & ... & ... & $2.88\pm0.19$ & ... & $0.72\pm0.32$ & $0.68\pm0.19$ & $12.66\pm0.60$ & ... & ... \\
05475 & 3.6060732 & -30.3801651 & 2.691 & $9.38\pm0.34$ & ... & $4.09\pm0.60$ & $5.15\pm0.39$ & $22.07\pm0.59$ & ... & ... & $2.06\pm0.19$ & $0.39\pm0.20$ & $2.03\pm0.22$ & $1.71\pm0.26$ & $5.77\pm0.41$ & ... & ... \\
05526 & 3.5914083 & -30.3797763 & 2.718 & $5.31\pm0.07$ & $1.63\pm0.09$ & $2.07\pm0.09$ & $3.63\pm0.10$ & $41.78\pm0.24$ & ... & ... & $3.65\pm0.25$ & $1.52\pm0.23$ & $1.19\pm0.28$ & $1.94\pm0.37$ & $23.44\pm0.73$ & ... & ... \\
06057 & 3.6033100 & -30.3742575 & 3.043 & $1.50\pm0.10$ & $0.51\pm0.17$ & ... & $1.04\pm0.22$ & $6.87\pm0.44$ & ... & ... & $0.57\pm0.08$ & $0.17\pm0.10$ & ... & $0.51\pm0.18$ & $2.90\pm0.38$ & ... & ... \\
\enddata
% - - - - - - - - - - - - - - - - - - - - - - - - - - - - - - - - - - - - - - - - - -
    \tablecomments{
    The first 4 columns are the same as Tab.\ref{tab:indi}.
    Columns 5-11 and 12-18 are the 2D/1D forward modeling flux, respectively for each emission line.
    The error bars shown in the table correspond to 1-$\sigma$ confidence intervals.
    %, whereas the upper/lower limits denote 2-$\sigma$ confidence limits. The table consists of four sections, corresponding to the three mass bins selected for stacking, and the AGN sample. The high, intermediate, and low mass bins are defined as $\log(\Mstar/\Msun) \in$~[10.0, 10.4), [9.7, 10.0), and [9.0, 9.7), respectively (see Sect.~\ref{subsect:stack}).}
    %\tablenotetext{a}{The superscript indicates the sets of strong line calibrations (B18: \citet{Bian:2018km}, C17: \citet{Curti:2017fn}) used in the metallicity inference, with the coefficients presented in Table~\ref{tab:coef}. We consider the results based on the B18 calibrations as our default results, for the sake of a direct comparison with the field measurements in \citet{Sanders:2021ga} derived using the same set of calibrations.}
    %\tablenotetext{b}{For sources in this section, their \oh estimates are not trustworthy since their nebular emissions are dominated by AGN ionization and therefore strong line calibrations are no longer applicable.
    }
% \output{\shipout\box255}
\end{deluxetable*}
% = = = = = = = = = = = = = = = = = = = = = = = = = = = = = = = = = = = = = = = = = =
}

\end{longrotatetable}
%= = = = = = = = = = = = = = = = = = = = = = = = = = = = = = = = = = = = = = = =


\bibliography{sample631}{}
\bibliographystyle{aasjournal}



\end{document}